\documentclass[8pt,a4paper,twocolumn]{article}

\usepackage[switch]{lineno} 
\usepackage{multicol,lipsum}

\usepackage{subfiles}

\usepackage{abstract}
\usepackage[table, dvipsnames]{xcolor}%Added for table coloring
\definecolor{lightgray}{gray}{0.85}

\usepackage[top=1.8cm,left=1.75cm,right=1.75cm,bottom=1.75cm]{geometry}
\usepackage[superscript,biblabel]{cite}
\usepackage{relsize}
\usepackage{tabularx}
\usepackage{array}
\usepackage{wrapfig}
\usepackage{multirow}
\usepackage{booktabs}
\usepackage{caption}
\usepackage{graphicx}
\usepackage{xcolor}
\usepackage{comment}

\usepackage[affil-it]{authblk} 
\usepackage{etoolbox}

\usepackage{lmodern}
\usepackage{amsmath}
\usepackage{mathtools}

\makeatletter
\patchcmd{\@maketitle}{\LARGE \@title}{\fontsize{16}{19.2}\selectfont\@title}{}{}
\makeatother

\begin{document}

\title{\textbf{{Longitudinal piezoelectric resonant photoelastic modulator for efficient intensity modulation at megahertz frequencies}}}

\author[1*]{Okan Atalar}
\author[2]{Rapha\"{e}l Van Laer}
\author[2]{Amir H. Safavi-Naeini}
\author[1]{Amin Arbabian}

\affil[1]{\textit{Department of Electrical Engineering, Stanford University, Stanford, California 94305, USA}}
\affil[2]{\textit{Department of Applied Physics and Ginzton Laboratory, Stanford University, Stanford, California 94305, USA}}
\affil[*]{Corresponding author: okan@stanford.edu\vspace{-2em}}

\date{}
\twocolumn[
  \begin{@twocolumnfalse}
\maketitle

\thispagestyle{empty}
%Having control over the properties of free-space beams is fundamental. Some properties include intensity, polarization, angle, .... SUch approaches have been used to manipulate the polarization, angle, intensity, respectively. 

%Manipulating the properties of an optical free-space beam is a fundamental problem in optics. 
%, and single frequency free-space intensity modulators (SFFIM), a special type of intensity modulator, modulate the intensity of a free-space beam at a fixed frequency.

%\textbf{The working principle of a longitudinal piezoelectric resonant photoelastic modulator operating at megahertz frequencies and functioning as a free-space intensity modulator is demonstrated. The free-space intensity modulator consists of a Y-cut lithium niobate wafer sandwiched between polarizers. The proposed free-space intensity modulator has low insertion loss, high acceptance angle (with passive polarization manipulating metasurface), and record breaking modulation efficiency that is more than an order of magnitude higher than other free-space intensity modulators operating in the megahertz frequency regime. The proposed modulator is compact and easily manufacturable. As a potential application, we  show that the proposed modulator can be integrated with a standard image sensor (e.g. CCD or CMOS) to effectively convert it into a time-of-flight imaging system.} 

\begin{abstract}
Intensity modulators are an essential component in optics for controlling free-space beams. Many applications require the intensity of a free-space beam to be modulated at a single frequency, including wide-field lock-in detection for sensitive measurements, mode-locking in lasers, and phase-shift time-of-flight imaging (LiDAR). Here, we report a new type of single frequency intensity modulator that we refer to as a longitudinal piezoelectric resonant photoelastic modulator. The modulator consists of a thin lithium niobate wafer coated with transparent surface electrodes. One of the fundamental acoustic modes of the modulator is excited through the surface electrodes, confining an acoustic standing wave to the electrode region. The modulator is placed between optical polarizers; light propagating through the modulator and polarizers is intensity modulated with a wide acceptance angle and record breaking modulation efficiency in the megahertz frequency regime. As an illustration of the potential of our approach, we show that the proposed modulator can be integrated with a standard image sensor to effectively convert it into a time-of-flight imaging system.\vspace{2em}
\end{abstract}

\end{@twocolumnfalse}
]

Controlling the intensity of a free-space optical beam is a problem of fundamental importance in optics. In this work, we focus on single frequency free-space intensity modulators (SFFIM), a special type of intensity modulator that modulates the intensity of a free-space beam at a fixed frequency. The application spaces for these modulators include wide-field lock-in detection for sensitive measurements~\cite{fluorescence_nature_imaging,video_rate_imaging_science,electro_optic_lifetime,resonant_electro_optic_lifetime,optical_lock_in_GW}, mode-locking in lasers~\cite{IR_free_space_ghz,graphene_EOM}, and phase-shift time-of-flight imaging (LiDAR)~\cite{miller2020large,electroabsorption_patent,optical_resonator,ToF_atalar}. It is desirable for an SFFIM to have low optical insertion loss, high modulation efficiency, and large acceptance angle~\cite{sun2016optical,reed2010silicon}. The modulation frequency is application specific, however, for many applications a higher frequency is desirable, especially for lock-in detection dominated by low frequency noise and time-of-flight (ToF) imaging, where the ranging accuracy is proportional to the frequency of intensity modulation.

 \begin{figure*}[t!]
\centering
\includegraphics[width=0.75\textwidth]{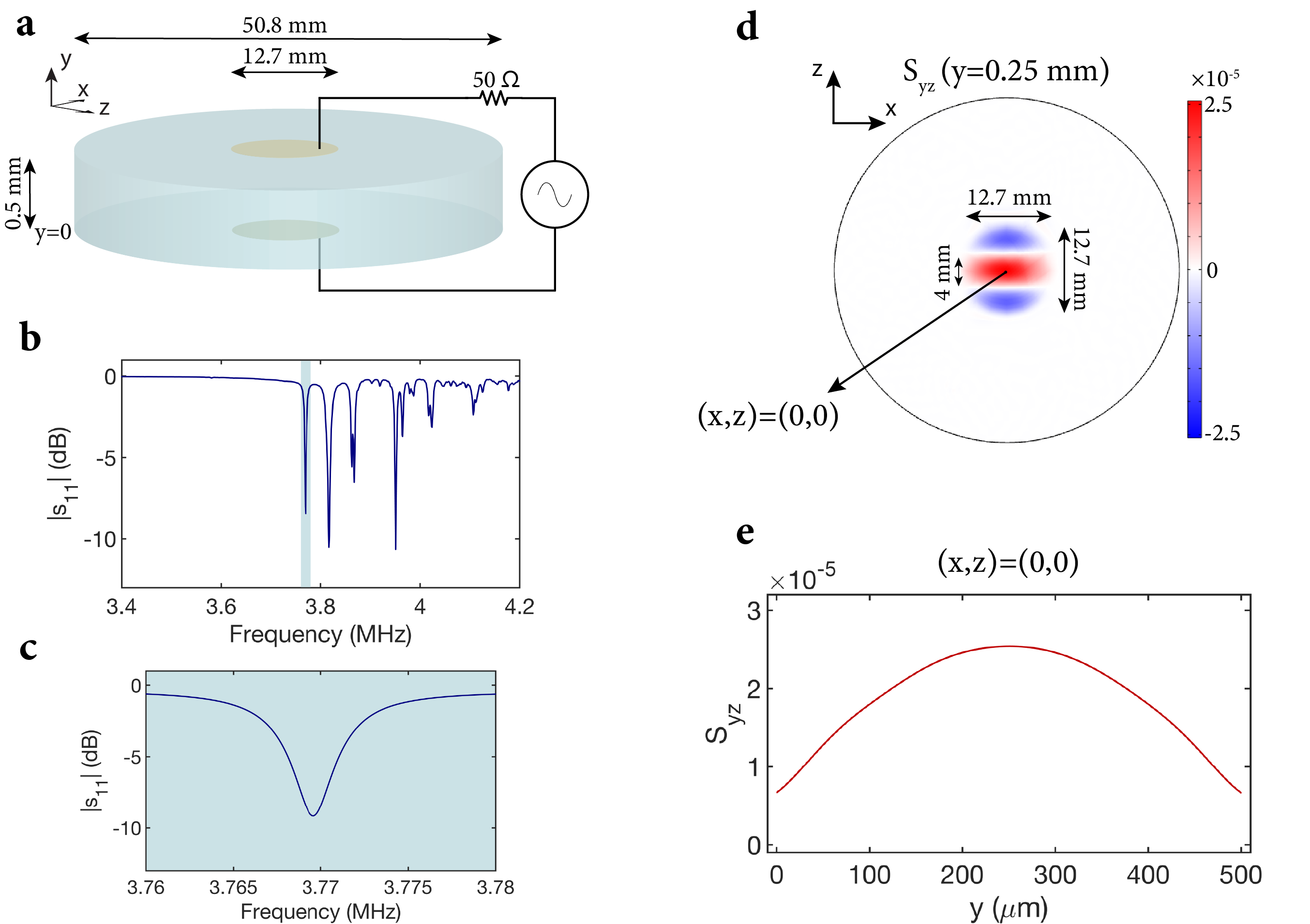}
\caption{\textbf{Exciting the correct acoustic mode of the wafer}. \textbf{a}, A Y-cut lithium niobate wafer of diameter 50.8~mm and of thickness 0.5~mm is coated on top and bottom surfaces with electrodes having a diameter of 12.7~mm. The wafer is excited with an RF source through the top and bottom electrodes. \textbf{b}, Simulated $|s_{11}|$ of the wafer with respect to 50~$\Omega$, showing the resonances corresponding to different acoustic modes of the wafer (loss was added to lithium niobate to make it consistent with experimental results). The desired acoustic mode appears around 3.77~MHz and is highlighted in blue. \textbf{c,} The desired acoustic mode $|s_{11}|$ with respect to 50~$\Omega$ is shown in more detail.
\textbf{d,} The dominant strain distribution ($S_{yz})$ when the wafer is excited at 3.7696~MHz with 2Vpp is shown for the center of the wafer. This strain distribution corresponds to the $|s_{11}|$ resonance shown in \textbf{c}. \textbf{e,} The variation in $S_{yz}$ parallel to the wafer normal and centered along the wafer is shown when the wafer is excited at 3.7696~MHz with 2Vpp.}
\label{fig:system1}
\end{figure*}

Achieving low optical insertion loss, high modulation efficiency, and large acceptance angle for an SFFIM at megahertz frequencies is a difficult and important challenge in optics. For modulation of light in the kilohertz frequency regime, liquid crystals offer good performance with low insertion losses and high modulation efficiencies~\cite{LC_1,LC_2,LC_3}. Pockels cells (longitudinal and transverse), electroabsorption modulators using the quantum-confined Stark effect, and Bragg cells based approaches allow modulation up to gigahertz frequencies. However, Bragg cells have narrow acceptance angle~\cite{double_pass_AOM}, electroabsorption modulators have limited modulation efficiencies and face challenges if scaled to centimeter square areas~\cite{electroabsorption_patent}, and Pockels cells having high modulation efficiencies need centimeter thick, bulky crystals~\cite{resonant_transverse_Pockels,longitudinal_Pockels}. These bulky modulators are highly sensitive to environmental conditions (e.g. temperature variations). Transverse resonant photoelastic modulators, where the acoustic wave propagates perpendicularly to the optical wave, have an inherent trade-off between aperture size and modulation frequency -- resulting in the aperture to be sub-millimeter for megahertz frequencies~\cite{transverse_photoelastic_mod}. This renders them unsuitable for many applications~\cite{fluorescence_nature_imaging,video_rate_imaging_science,electro_optic_lifetime,resonant_electro_optic_lifetime,optical_lock_in_GW,miller2020large,electroabsorption_patent,optical_resonator,ToF_atalar}. Spatial light modulators (SLMs) infused with liquid crystals lack the speed to be operated at megahertz frequencies, and new types of high-speed SLMs have limited modulation efficiencies~\cite{reflection_control_metasurface,slm_steering_brongersma,gate_tunable_metasurface}. Metasurfaces have made significant progress over the last decade and have allowed unprecedented control over the properties of free-space beams~\cite{broadband_metalens,metasurface_polarization_transf,ultra_high_q_plasmonic_metasurface}. However, active metasurfaces are still in their infancy and have limited functionality~\cite{spatiotemporal_review}. A better solution for free-space optical modulation is urgently needed.

\begin{figure*}[t!]
\centering
\includegraphics[width=0.75\textwidth]{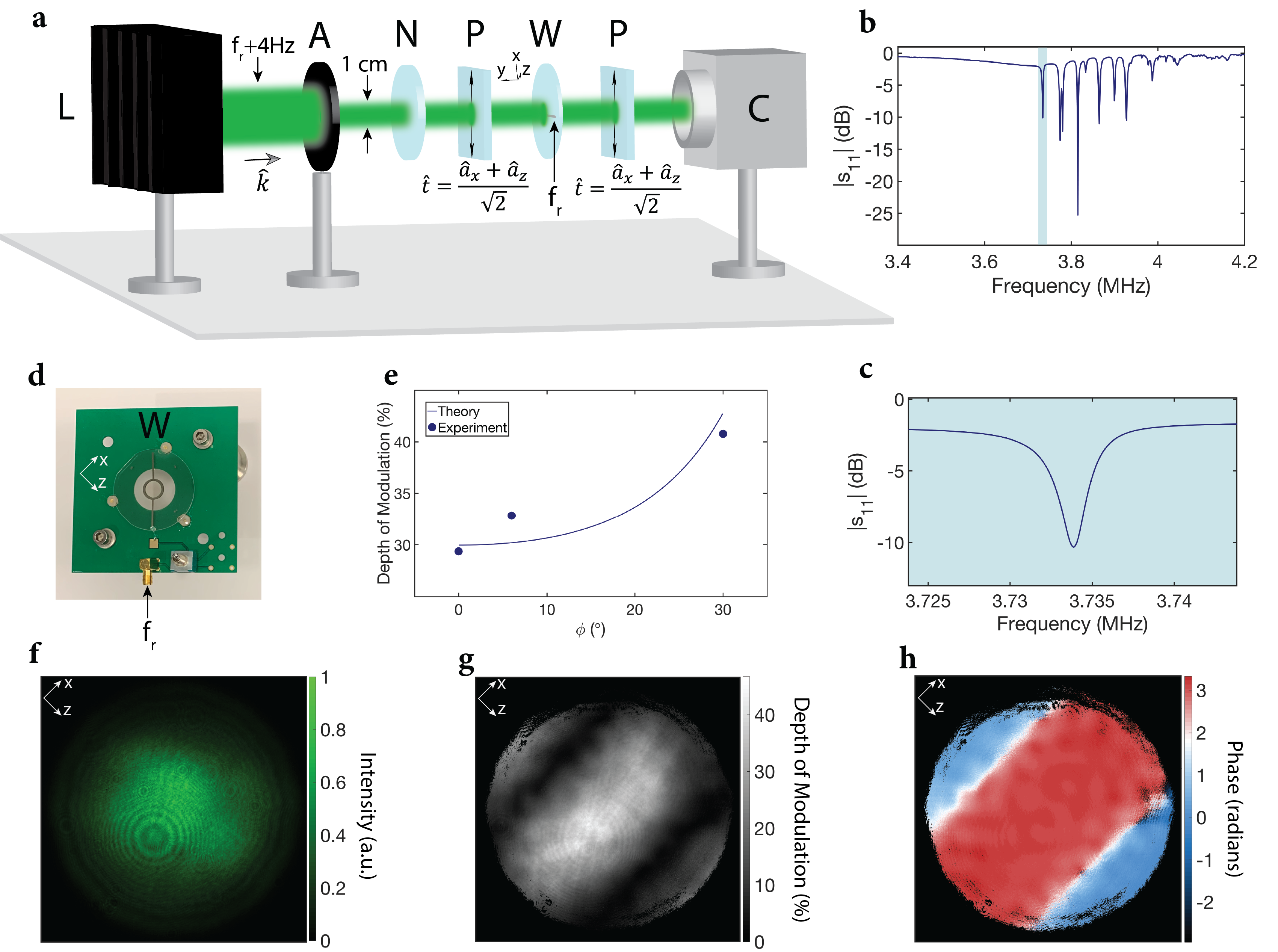}
\caption{\textbf{Experimental characterization of the modulator}. \textbf{a}, Schematic of the characterization setup is shown. The setup includes a laser (L) with a wavelength of 532~nm that is intensity modulated at 3.733704~MHz, aperture (A) with a diameter of 1~cm, neutral density filter (N), two polarizers (P) with transmission axis $\hat{t}$, wafer (W), and a standard CMOS camera (C). The wafer is excited with 90~mW of RF power at $f_r = 3.7337$~MHz, and the laser beam passes through the center of the wafer that is coated with ITO. The camera detects the intensity modulated laser beam. \textbf{b}, The desired acoustic mode is found for the modulator by performing an $s_{11}$ scan with respect to 50~$\Omega$ using 0~dBm excitation power and with a bandwidth of 100~Hz. The desired acoustic mode is highlighted in blue.
\textbf{c,} The desired acoustic mode is shown in more detail by performing an $s_{11}$ scan with respect to 50~$\Omega$ using 0~dBm excitation power with a bandwidth of 20~Hz. \textbf{d,} The fabricated modulator is shown. \textbf{e}, The depth of intensity modulation is plotted for different angles of incidence for the laser beam, where $\phi$ is the angle between the surface normal of the wafer and the beam direction $\hat{k}$ (see Methods section for more details). \textbf{f,} Time-averaged intensity profile of the laser beam detected by the camera is shown for $\phi = 0$. \textbf{g,} The DoM at 4~Hz of the laser beam is shown per pixel for $\phi = 0$. \textbf{h,} The phase of intensity modulation at 4~Hz of the laser beam is shown per pixel for $\phi = 0$.}
\label{fig:system2}
\end{figure*}

%with potentially high acceptance angle (if coated with a specially designed static polarization manipulating metasurface, an area which has seen significant progress recently).

%going over an order of magnitude beyond

In this paper, we demonstrate a new type of intensity modulator that is compact, easy to manufacture, requires no DC bias and has record modulation efficiency at megahertz frequencies. We demonstrate efficient intensity modulation at 3.7~MHz over a cm$^{2}$ scale area, going beyond any other free-space intensity modulator. We also demonstrate efficient modulation for multiple angles of incidence. Illustrating the strength of our system, we experimentally show that the modulator can be used with a standard camera to enable high spatial resolution ToF imaging. Our approach enables an attractive alternative to complex LiDAR systems relying on optical phased arrays (OPAs) and SLMs~\cite{SiN_wide_angle_beam_steering,parallel_steering_nature,time_stretch_lidar_nature,serpentine_OPA,slm_steering_science,slm_steering_brongersma} with thousands of control elements or highly specialized costly image sensors that are difficult to implement with a large number of pixels~\cite{megapixel_spad_optica,SPAD_IEEE,ToF_PMD_IEEE}.

%\section*{Time-of-Flight Imaging Using the Modulator}
\begin{figure*}[t!]
\centering
\includegraphics[width=0.8\textwidth]{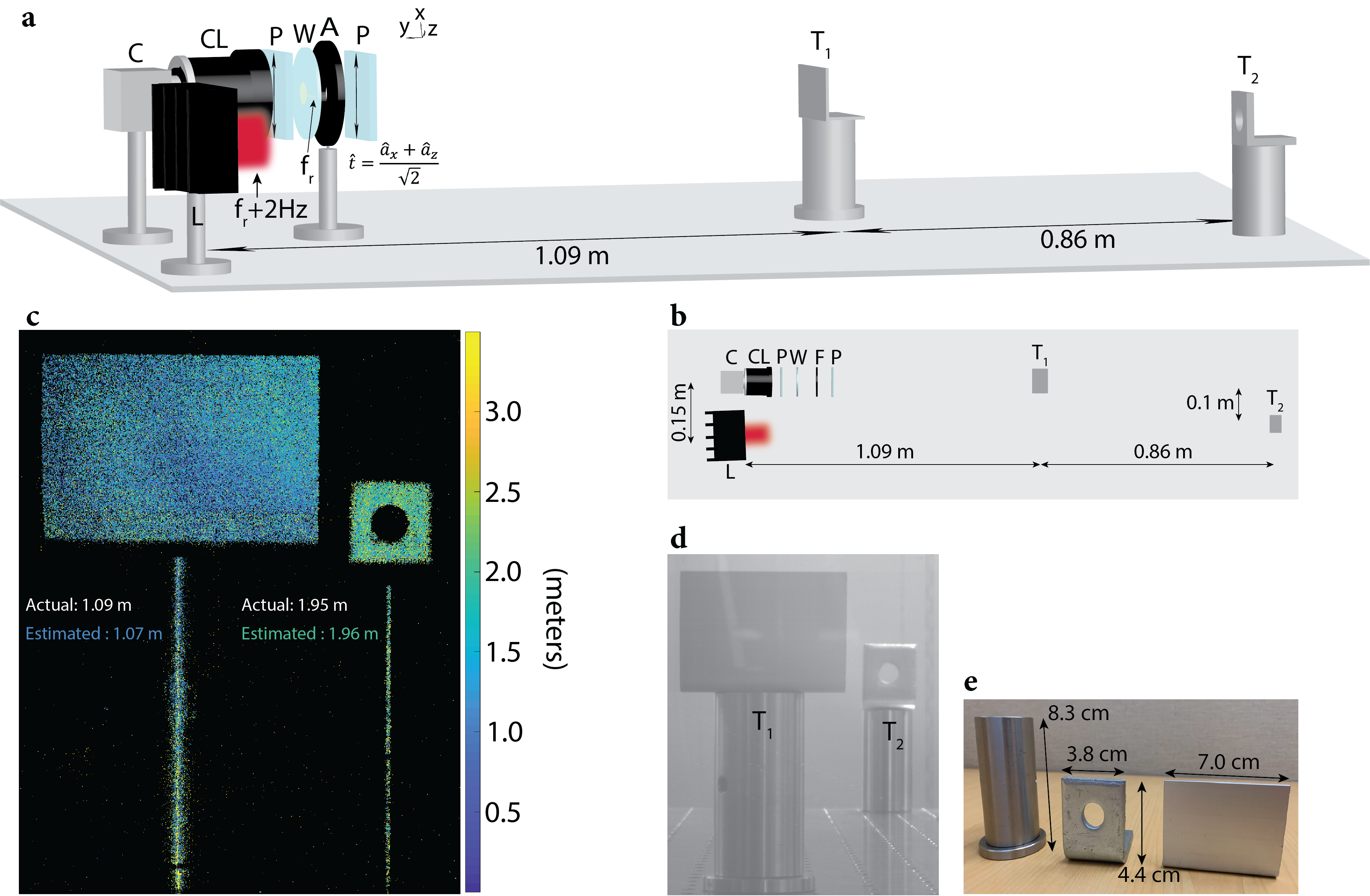}
\caption{\textbf{ToF imaging using the modulator and a standard camera}. \textbf{a}, Schematic of the imaging setup is shown. The setup includes a standard CMOS camera (C), camera lens (CL), two polarizers (P) with transmission axis $\hat{t}$, wafer (W), aperture (A) with a diameter of 4~mm, laser (L) with a wavelength of 635~nm that is intensity modulated at 3.733702~MHz, and two metallic targets ($T_1$ and $T_2$) placed 1.09~m and 1.95~m away from the imaging system, respectively. 140~mW of RF power at $f_r = 3.7337$~MHz is used to excite the wafer electrodes. The laser is used for illuminating the targets. The camera detects the reflected laser beam from the two targets, and uses the 2~Hz beat tone to extract the distance of each pixel corresponding to a distinct point in the scene (see Methods section for more details). \textbf{b}, Bird's eye view of the schematic in \textbf{a}.
\textbf{c,} Reconstructed depth map seen by the camera. Reconstruction is performed by mapping the phase of the beat tone at 2~Hz to distance using equation \eqref{Eq.R4}. The distance of each pixel is color coded from 0 to 3 meters (pixels that receive very few photons are displayed in black). The distance of targets $T_1$ and $T_2$ are estimated by averaging across their corresponding pixels, respectively. The estimated distances for $T_1$ and $T_2$ are 1.07~m and 1.96~m, respectively. \textbf{d,} Ambient image capture of the field-of-view of the camera, showing the two targets $T_1$ and $T_2$. \textbf{e}, The dimensions of the targets used for ToF imaging are shown.}
\label{fig:system3}
\end{figure*}

Resonant designs are commonly used to improve the modulation efficiency of optical modulators operating at a single frequency. However, optical resonators with high quality factors come at the expense of a limited acceptance angle, while the attainable quality factors are limited for radio frequency (RF) resonators. To achieve both a high quality factor and to break the trade-off between quality factor and acceptance angle (and therefore modulation efficiency) we choose to use an acoustic resonator. To construct an efficient intensity modulator incorporating an acoustic resonator and that is able to couple RF to optics, a material with suitable piezoelectric and photoelastic properties should be chosen. Here we use lithium niobate (LN) as the modulator material, since it offers good piezoelectric and photoelastic properties, and is widely available at low-cost.

We use a Y-cut LN wafer to break in-plane symmetry and allow the x and z directions to be modulated differently when an electric field is applied along the y direction of the wafer. To reach megahertz modulation frequencies over a centimeter square area, the thickness of the wafer needs to be chosen such that the fundamental acoustic mode appears around megahertz frequencies. To satisfy this requirement, we use an LN wafer that is 0.5~mm thick with a diameter of 50.8~mm. The top and bottom surfaces of the wafer are coated with electrodes with a diameter of 12.7~mm. This allows us to reach a centimeter square scale input aperture, while simultaneously confining the acoustic mode to the electrode region to limit clamping losses. The modulator is shown in Fig.~\ref{fig:system1}\textbf{a}. The modulator has many acoustic modes that can be excited. These modes are shown in Fig.~\ref{fig:system1}\textbf{b}, obtained by simulating the $|s_{11}|$ of the LN wafer using COMSOL~\cite{COMSOL5}.

\begin{table*}[ht]
\centering 
    \begin{tabular}{ | l | l | p{2.95cm} | p{3.65cm}| p{1.9cm} |}
    \hline \rowcolor{lightgray}
    \textbf{Modulation principle} & \textbf{$\lambda$ (nm)} & \textbf{Optical insertion loss (dB)} & \textbf{RF power required to achieve 100\% intensity modulation (W/$\text{cm}^{2}$)} & \textbf{Optically broadband} 
    \\ \hline
    Transverse resonant Pockels effect~\cite{resonant_transverse_Pockels} & 532 & 3 & 6.7 $\times 10^5$ & Yes \\ \hline
    Longitudinal Pockels effect~\cite{longitudinal_Pockels} & 532 & 3 & 1.83 $\times 10^4$ & Yes \\
    \hline
    Electroabsorption~\cite{electroabsorption_patent} & 860 & - & 34 & No \\ \hline
    Gate-tunable metasurface~\cite{gate_tunable_metasurface} & 1,550 & - & 460 & Yes \\ \hline
    Plasmonic nanoresonator~\cite{slm_steering_brongersma} & 1,340 & 20 & 454 & Yes
    \\ \hline
    \textbf{Photoelastic modulation*} & \textbf{532} & \textbf{4.2} & \textbf{7.4} & \textbf{Yes} \\ \hline
    \end{tabular}
    %\label{fig:system}
    %caption{\label{tab:table-name}Your caption.}
    %\caption*{The caption without a number}
    \caption{\textbf{Comparison table}. Modulator performances are compared in terms of intensity modulation at 3.7~MHz. The two performance metrics used are the optical insertion loss and the RF power required to achieve 100\% intensity modulation of the laser beam. To show modulation efficiency, the RF power required to achieve 100\% intensity modulation is calculated by extrapolating the values reported in the references.}
\label{table:1}
    \end{table*}

The ideal acoustic mode has large piezoelectric and photoelastic couplings, as well as a uniform strain distribution in amplitude and phase in the electrode region. One such mode is the fundamental one shown in Fig.~\ref{fig:system1}\textbf{c} - in fact, this is the only acoustic mode that can be used for this modulator to achieve fairly uniform and efficient modulation. We simulate the effect of a 2Vpp signal applied to the surface electrodes using COMSOL at $f_c = 3.7696~\text{MHz}$ (corresponding to the resonance in $|s_{11}|$ for Fig.~\ref{fig:system1}\textbf{c}). The dominant strain $S_{yz}$ is shown in Fig.~\ref{fig:system1}\textbf{d}. The contribution of the other strain components is negligible compared to $S_{yz}$ (see Supplementary Information Section 1 for details). Since an acoustic standing wave is excited, the phase of the strain is 0 or $\pi$ radians. The acoustic mode in Fig.~\ref{fig:system1}\textbf{d} is composed of three different regions with roughly uniform strain amplitude, but with different phases (displayed as red and blue regions in Fig.~\ref{fig:system1}\textbf{d}). 

The strain in the wafer causes time-varying birefringence at the frequency $f_c$ of the signal applied to the electrodes. This is expressed in equation \eqref{Eq.R1}, where $n_o$ and $n_e$ are the ordinary and extraordinary refractive indices of LN, $p_{14}$ is the photoelastic constant of LN which relates the volume average shear strain ${\bar{S}_{yz} = A \text{cos}(2 \pi f_c t)}$ to the index ellipsoid of LN. $n_x(t)$, $n_y(t)$, and $n_z(t)$ are the modulated refractive indices of the LN wafer. The volume average shear strain magnitude is $A = 1.13 \times 10^{-5}$ for the COMSOL simulation. This volume average strain is calculated for a 1~cm diameter region centered on the wafer and over a thickness of 0.5~mm. 

\begin{gather}
\frac{1}{n_x^2(t)} = \frac{1}{n_o^2} + 2p_{14}\bar{S}_{yz} \nonumber \\
\frac{1}{n_y^2(t)} = \frac{1}{n_o^2} - 2p_{14}\bar{S}_{yz} \nonumber \\
\frac{1}{n_z^2(t)} = \frac{1}{n_e^2} 
\label{Eq.R1}
\end{gather}

A laser beam propagating through the transparent electrode region excites an ordinary and an extraordinary wave in the wafer (due to anisotropy of LN), and these two waves pick up time-varying phases $\phi_o(t)$ and $\phi_e(t)$ upon propagation through the wafer, respectively. This results in the polarization of the laser beam that has propagated through the wafer to rotate in time with frequency $f_c$. 

%A laser beam passing through the wafer electrode region can be treated as the linear superposition of plane waves with different angles of incidence. The modulation of the laser beam can be computed by finding the modulation for each plane wave. A plane wave propagating through the electrode region excites an ordinary and an extraordinary wave in the wafer (due to anisotropy of LN), and these two waves pick up time-varying phases $\phi_o(t)$ and $\phi_e(t)$ upon propagation through the wafer, respectively. This leads to the polarization of the plane wave that has passed through the wafer to rotate in time with frequency $f^*$. 

%These two waves pick up phases $phi_o(t)$ and $phi_e(t)$ after propagating through the wafer, respectively (see Supplementary Information for details). 

%\begin{gather}
%\phi_o(t) = \frac{2 \pi L}{\lambda}n_o(t)cos(\tilde{\theta}_o) \nonumber \\
%\phi_e(t) = \frac{2 \pi L}{\lambda}n_e(t)cos(\tilde{\theta}_e) \label{Eq.2}
%\end{gather}

%Since the ordinary and extraordinary waves pick up different time-varying phases upon propagating through the wafer, the output polarization of the plane wave rotates in time at the applied RF frequency $f_n$. The output polarization of the plane wave that has traversed the wafer. 

%\begin{gather}
%\bar{p}_f(t) = c_o\bar{p}_{oi}\text{exp}(j\phi_o(t)) + c_e\bar{p}_{ei}\text{exp}(j\phi_e(t)) \label{Eq.3}
%\end{gather}

We place the wafer between optical polarizers with transmission axis $\hat{t} = (\hat{a}_x + \hat{a}_z)/\sqrt{2}$ to convert polarization modulation into intensity modulation. The intensity $I(t)$ of a plane wave with random polarization that has passed through the three components is expressed in equation \eqref{Eq.R2} (see Supplementary Information Section 2 for derivation), where $I_0$ is the intensity of the incoming plane wave, $c_o$ is the amplitude of the excited ordinary wave, $c_e$ is the amplitude of the excited extraordinary wave, HOH stands for the higher order harmonics, $J_0$ and $J_1$ stand for the zeroth and first order Bessel functions of the first kind, respectively. The static phase accumulation $\phi_s$ and the dynamic phase accumulation $\phi_D$ are found as follows: $\phi_s + \phi_D \text{cos}(2 \pi f_c t) = \phi_o(t) - \phi_e(t)$. 

%The device we report in this work bears resemblance to an acousto-optic tunable filter relying on anisotropic Bragg diffraction (reference). However, the modulation mechanism of light are different for these two approaches.    

%\begin{gather}
%I(t) = \frac{I_0}{2(c_o^4 + 2c_o^2c_e^2 + c_e^4)}(c_o^4 + c_e^4 + 2c_o^2c_e^2[\text{cos}(\phi_s)(J_o(\phi_D) \nonumber \\ - 2\text{sin}(\phi_s)J_{1}(\phi_D)\text{cos}(2 \pi f^* t) + \text{HOH}]) \label{Eq.2}
%\end{gather}

\begin{gather}
I(t) = \frac{I_0}{2}\Big(c_o^4 + c_e^4 + 2c_o^2c_e^2\big[\text{cos}(\phi_s)(J_o(\phi_D) \nonumber \\ - 2\text{sin}(\phi_s)J_{1}(\phi_D)\text{cos}(2 \pi f_c t) + \text{HOH}\big]\Big) \label{Eq.R2}
\end{gather}

To demonstrate intensity modulation using this approach, we coat the top and bottom surfaces of a double side polished 50.8~mm diameter and 0.5~mm thick Y-cut LN wafer with indium tin oxide (ITO) to serve as transparent surface electrodes. The desired acoustic mode is found by measuring the reflection scattering parameter $s_{11}$ using a vector network analyzer (VNA). The dips in $|s_{11}|$ occur at similar frequencies as in the COMSOL simulation shown in Fig.~\ref{fig:system1}\textbf{c}, with the fundamental appearing around 3.73~MHz. The measured $|s_{11}|$ for the desired acoustic mode of the wafer is shown in Fig.~\ref{fig:system2}\textbf{c}, with a quality factor of around $10^3$. To measure the intensity modulation efficiency and to verify that the acoustic mode profile matches simulation (Fig.~\ref{fig:system1}\textbf{d}), the coated wafer is placed between optical polarizers and a laser beam with wavelength 532~nm is passed through the polarizer, wafer electrode region, and polarizer, as shown in Fig.~\ref{fig:system2}\textbf{a}. 90~mW of RF power is applied to the surface electrodes at $f_r = 3.7337~\text{MHz}$, which corresponds to the resonance observed in $|s_{11}|$ in Fig.~\ref{fig:system2}\textbf{c}. 90~mW of RF power corresponds to 6Vpp over the wafer surface electrodes.

To detect the intensity modulation profile of the laser beam with high spatial resolution, we use a standard CMOS camera offering four megapixel resolution. Since the frame rate of the camera is limited to less than a kilohertz, and the modulator operates at $f_r = 3.7337~\text{MHz}$, we  intensity modulate the laser beam at $f_r + 4~\text{Hz}$ by using a free-space acousto-optic modulator (see Methods for details) to perform heterodyne detection and measure the 4~Hz beat tone using the camera, as shown in Fig.~\ref{fig:system2}\textbf{a}. The depth of intensity modulation is proportional to the amplitude of the beat tone, and its phase is related to the acoustic standing wave phase. The laser beam intensity profile, the normalized amplitude of the beat tone (depth of modulation), and the phase of the beat tone are shown in Fig.~\ref{fig:system2}\textbf{f,g,h}, respectively. The inferred volume average strain over 1~cm diameter region centered on the wafer and over a thickness of 0.5~mm for the experiment is $4.15 \times 10^{-5}$ (see Supplementary Information Section 4 for details). This agrees well with the COMSOL simulation ($1.13 \times 10^{-5} \times \frac{6\text{Vpp}}{2\text{Vpp}} = 3.39 \times 10^{-5}$). The ability to modulate different angles of incidence is shown in Fig.~\ref{fig:system2}\textbf{e}, where $\phi$ is the angle between the wafer normal and the laser beam propagation direction (see Methods for more details). 

%\begin{gather}
%\text{DoM} = \frac{4c_o^2 c_e^2 J_1(\phi_D) \text{sin}\phi_s }{c_o^4 + c_e^4 + 2c_o^2 c_e^2 J_0(\phi_D) \text{cos}\phi_s } \label{Eq.3}
%\end{gather}

We use the fabricated modulator for phase-shift based ToF imaging~\cite{ToF_PMD_IEEE}. In this imaging modality, intensity modulated light is used to illuminate targets in a scene. The targets at different distances to the transmitter reflect back to the receiver with different phase shifts, where the phase shift $\Phi$ in radians is related to the distance $d$, speed of light in air $c$, and the intensity modulation frequency $f_m$ as shown in equation \eqref{Eq.R4}. 

\begin{gather}
d = \frac{\Phi c}{4 \pi f_m} \label{Eq.R4}
\end{gather}

Higher modulation frequencies are favorable due to providing better range resolution. Megahertz frequencies are usually used since the unambiguous imaging range of $c/(2 f_m)$ due to phase wrapping is on the meters level, while also offering good range resolution. Since standard cameras have frame rates limited to less than kilohertz frequencies, they cannot sample the megahertz modulation frequency. The modulator allows the phase of the megahertz signal to be detected by the camera through heterodyne detection.

The ToF imaging setup using a standard camera and the modulator is shown in Fig.~\ref{fig:system3}\textbf{a}. A laser of wavelength 635~nm is intensity modulated at 3.733702~MHz and is used to illuminate two targets. We place the modulator between two optical polarizers, and we place an aperture in front of the wafer to only use the center 4~mm diameter region so that destructive interference between the anti-phase regions is avoided. We attach a camera lens to the camera to resolve the two targets. Applying equation \eqref{Eq.R4} to the captured frames having a beat tone at 2~Hz, the distance that each camera pixel corresponds to is calculated, and the reconstructed depth map is shown in Fig.~\ref{fig:system3}\textbf{c}. The target distances are reconstructed to centimeter level accuracy. The reason why only a thin slice of the optical posts are seen in the depth reconstruction for Fig.~\ref{fig:system3}\textbf{c} is due to specular reflection from the shiny metal posts. This is a common problem for all LiDAR and optical ToF imaging systems~\cite{specular_1,specular_2}.

The experiment shows that high spatial resolution ToF imaging can be achieved with a standard camera and the modulator. It also demonstrates the optically broadband nature of the modulator, being able to modulate multiple different wavelengths (limited by the transparency window of LN and ITO). The high spatial resolution for the depth reconstruction is enabled by the megapixel resolution of the camera. The imaging performance could be significantly improved by depositing a static polarization manipulating metasurface~\cite{metasurface_polarization_static} on the electrodes of the wafer. Such a metasurface could allow the whole aperture to be utilized (compensating for the phase of the anti-phase regions shown in Fig.~\ref{fig:system1}\textbf{d}), and also allow the modulation efficiency to remain high for a large acceptance range by canceling the static phase variation $\phi_s$ for different angles of incidence.

A comparison table is provided in Table 1 for different modulation approaches. The two performance metrics used are optical insertion loss and the RF power required to achieve 100\% intensity modulation of the laser beam. Only modulators that can achieve large acceptance angles are included. For the Pockels cell approaches, a thickness of 0.5~mm along the light propagation direction is assumed to have the same temperature tolerance as reported in this work. The 4.2~dB optical insertion loss reported could be brought closer to 3~dB (due to polarizers) by depositing anti-reflection coatings on the wafer. We have neglected cut-off frequency issues due to capacitance and have only focused on efficiency. The modulators compared to in Table 1 are capacitive and will face challenges when scaled to centimeter square areas without matching circuits. The impedance of our modulator is well matched to 50~$\Omega$ and therefore does not require a matching circuit. In fact, the $7.4~\text{W}/\text{cm}^2$ reported is a worst case for our device, since more energy is required to achieve 100\% modulation due to the Bessel function compared to other modulation approaches (see Supplementary Information Section 6 for more details). The modulation efficiency of our approach could be further improved by investigating hybrid platforms including multiple layers of materials to separate the piezoelectric and photoelastic functionalities to different materials. These platforms could offer higher acoustic quality factors for some acoustic modes and therefore improve the modulation efficiency.

%We would also like to point out that our modulator outperforms other modulators in these metrics, while also offering an easier fabrication process.

To conclude, we have demonstrated a new resonant free-space intensity modulator that modulates light from visible and up to near infrared wavelengths at megahertz frequencies with record efficiency. The modulator can find immediate use in applications requiring free-space beams to be intensity modulated with low RF power at megahertz frequencies over centimeter square scale apertures. It could enable low-cost and high spatial resolution ToF imaging and LiDAR with low-cost standard image sensors.

\bibliographystyle{unsrt}
\bibliography{references}

\section*{Methods}
\subsection*{Modulator fabrication}

We fabricate the piezoelectric resonant photoelastic modulator using a 50.8~mm diameter, 0.5~mm thick double side polished Y-cut LN wafer. A metal plate with dimensions 63.5~mm by 63.5~mm and thickness 1~mm is drilled in the center with a hole diameter of 12.7~mm. The metal plate is centered and placed on top of the wafer and approximately 450~nm thick ITO is deposited using sputter coating in a load locked chamber. This sputtering process using the metal plate is repeated for the backside of the wafer. During the sputtering process, the chamber is not heated to prevent the formation of black LN. The optical transmission of ITO coated on the LN wafer is improved by heating on a hot plate for 26~minutes at a temperature of $250^\circ C$. The sheet resistance of ITO after heating is approximately $25~\Omega/\text{sq}$.

A metal plate with dimensions 63.5~mm by 63.5~mm and thickness 1~mm is drilled such that the metal region that lies between two circles centered on the plate with diameter 10.16~mm and 12.7~mm is removed. A rectangular region that is 6.35~mm away from the center of the plate and with dimensions 1~mm by 16.5~mm is also drilled. The metal plate is centered and placed on top of the wafer and 300~nm thick aluminum is evaporated in a load locked chamber. This evaporation process using the metal plate is repeated for the backside of the wafer. 

The ITO and aluminum coated wafer is attached to a PCB by gluing using epoxy to three plastic washers. The top surface electrode is electrically connected to the PCB signal port using two gold wirebonds, and the bottom surface electrode is electrically connected to the PCB ground using two gold wirebonds. The wirebonds for the top and bottom surfaces of the wafer connect to the microstrip aluminum region defined by the rectangular region (dimensions 1~mm by 16.5~mm) drilled on the metal plate.

\subsection*{Experimental setup for modulator characterization}
For RF characterization of the modulator, we perform an $s_{11}$ scan with respect to $50~\Omega$ using a VNA (Rohde\&Schwarz ZNB20) with excitation power of 0~dBm and bandwidth of 100~Hz for the broad scan with a frequency step size of 100~Hz (Fig.~\ref{fig:system2}\textbf{b}) and a bandwidth of 20~Hz and frequency step size of 10~Hz for the focused scan (Fig.~\ref{fig:system2}\textbf{c}), respectively. The modulator is excited through the SMA connector attached to the PCB holding the modulator (Fig.~\ref{fig:system2}\textbf{d}).

For optical characterization of the modulator, we use the setup shown in Fig.~\ref{fig:system2}\textbf{a}. In this setup, a diode-pumped solid-state (DPSS) laser diode  of wavelength 532~nm (Thorlabs DJ532-10) mounted on a laser mount (Thorlabs TCLDM9) is used. The laser beam is intensity modulated using a free-space acousto-optic modulator (G\&H AOMO 3080-125). The acousto-optic modulator (AOM) is excited at its center frequency of 80~MHz, and this carrier frequency is intensity modulated at 3.733704~MHz. This causes the light beam passing through the AOM to have diffracted beams, and these beams to be intensity modulated at 3.733704~MHz. We use the first order beam for characterizing the modulator. This beam is passed through an iris (Thorlabs ID25) with diameter adjusted to 1~cm to match the beam diameter to the active region of the modulator (ITO covered region). 

The intensity modulated laser beam passes through a variable neutral density filter (OptoSigma NDHN), a wire-grid polarizer (Thorlabs WP25L-VIS), the center of the modulator (the modulator is excited with 90~mW of power at 3.7337~MHz through the SMA connector of the PCB it is attached to), and then through another wire-grid polarizer. The signal generators driving the AOM and the proposed modulator are synchronized. The two polarizers and the wafer surfaces are placed parallel to each other and the centers of the three components are aligned. The wafer optic axis makes a $45^\circ$ angle with the transmission axis of the two polarizers. The laser beam that has passed through the first polarizer, the wafer, and the second polarizer (analyzer) is detected with a standard camera (Basler acA2040-90um) using a frame rate of 30~Hz, 16 bit precision per pixel, 400~$\mu s$ exposure time; 600 frames are captured. The depth of intensity modulation is calculated by performing a time-domain fast Fourier transform (FFT) on each pixel of the camera, and finding the ratio of the beat tone at 4~Hz multiplied by four to the DC level. The phase of modulation is calculated by finding the phase of the beat tone at 4~Hz after performing the FFT on each pixel. 

The intensity modulation using the AOM does not result in a homogeneous modulation of the laser beam. Since the acoustic wave is turned on and off at a frequency of 3.733704~MHz, the phase of the intensity modulation for the laser beam has a linear phase variation depending on the laser beam size and the acoustic wavelength corresponding to 3.733704~MHz in the AOM. This linear phase variation of the intensity modulation frequency of 3.733704~MHz for the laser beam varies along the direction of acoustic wave propagation in the AOM. This phase variation is removed in Fig.~\ref{fig:system2}\textbf{h} by a least-squares fit to the phase of modulation calculated after performing the FFT. 

Depth of intensity modulation for different angles of incidence of the laser beam to the wafer surface is also captured. The angle of incidence of the laser beam is such that it makes an equal angle with the $x$ and $z$ axes of the LN wafer. In standard spherical coordinate notation, the two angles used to describe the laser beam direction $\hat{k}$ can be expressed as $\theta = \text{cos}^{-1}\Big(\frac{\text{sin} \phi}{\sqrt{2}}\Big)$  and $\psi = \text{sin}^{-1}\Bigg(\frac{\text{cos} \phi}{\text{sin}\bigg(\text{cos}^{-1}\big(\frac{\text{sin} \phi}{\sqrt{2}}\big)\bigg)}\Bigg)$. The depth of intensity modulation is measured as described in the previous paragraph for each of the different angles of incidence. The values reported in Fig.~\ref{fig:system2}\textbf{e} are the mean DoM values of the pixels in the center region lying between the two null regions.

The optical insertion loss of the wafer is determined by passing the 532~nm laser beam through the ITO coated region of the wafer and measuring the DC level with a photodetector (Thorlabs DET36A2). The optical insertion loss is found to be 1.2~dB, calculated by taking the ratio of the light intensity measured by the photodetector with and without the wafer placed. The optical insertion loss of 4.2~dB reported in Table 1 also includes the 3~dB loss of the polarizer when unpolarized light is assumed. 

\subsection*{Experimental setup for ToF imaging}
A laser diode of wavelength 635~nm (Thorlabs HL6322G) and with output optical power less than 15~mW is mounted on a laser mount (Thorlabs TCLDM9) and intensity modulated at 3.733702~MHz using a signal generator. The laser diode has a divergence of $8^\circ$ by $30^\circ$ (FWHM). The depth of intensity modulation is approximately 100\% for the laser diode, measured with a photodetector (Thorlabs DET36A2). The intensity modulated laser beam is used to illuminate targets $T_1$ and $T_2$ shown in Fig.~\ref{fig:system3}. The reason why the modulator characterization is done using a DPSS laser with a wavelength of 532~nm is because of the stability and low divergence angle of the laser beam. The stable laser beam with low divergence angle from the DPSS laser allows the optical characterization to be performed. Modulating the DPSS laser through current modulation is limited to low frequencies (below 1~MHz). That is why an AOM was used to modulate the laser beam emitted by the DPSS laser. Intensity modulation using an AOM is avoided for ToF imaging due to the linear phase variation of the intensity modulation frequency across the laser beam (as described in the previous section). Since distance is encoded in the phase of the beat frequency, using an AOM for intensity modulation would make the depth map reconstruction difficult. This is the reason why a laser diode with a wavelength of 635~nm and larger divergence angle was used for performing the ToF imaging experiment.

For the ToF imaging setup, we place the wafer between two optical polarizers such that the transmission axis of the two polarizers makes a $45^\circ$ angle with the optical axis of the wafer, and the wafer and polarizer surfaces are placed parallel to each other. An iris (Thorlabs ID25) with diameter adjusted to 4~mm is placed between the polarizer and wafer such that the laser beam only passes through the center 4~mm of the wafer. This causes the laser beam to be approximately uniformly modulated by preventing the beam from being modulated by the anti-phase regions (see Fig.~\ref{fig:system1}\textbf{d}). The modulator is excited with 140~mW of power at 3.7337~MHz through the SMA connector of the PCB. The four components (polarizer, modulator, iris, polarizer) are placed in front of a standard camera (Basler acA2040-90um). The signal generators driving the laser diode and the modulator are synchronized. 

The reflected light from the two targets is captured by the camera. A camera lens with a diameter of 45.0~mm (Kowa LM75HC) is attached to the camera to resolve the two targets on the camera image sensor. The camera is operated with a frame rate of 10~Hz, 16 bit precision per pixel, exposure time of 99~ms, and internal gain of 23~dB; 600 frames are captured. The distance that each pixel corresponds to is calculated using equation \eqref{Eq.R4}. The resulting depth map is shown in Fig.~\ref{fig:system3}\textbf{c}. The standard deviation for the depth map in Fig.~\ref{fig:system3}\textbf{c} is 1.0~m for the pixels corresponding to $T_1$, and 0.92~m for the pixels corresponding to $T_2$. For capturing the ambient image (Fig.~\ref{fig:system3}\textbf{d}), 16 bit precision is used with 1~second of capture time and 11~dB internal gain for the camera. 

\section*{Acknowledgements}
The authors thank Christopher J. Sarabalis for useful discussions. This work was funded in part by Stanford SystemX Alliance, Office of Naval Research, and NSF ECCS-1808100.

\section*{Author contributions}
O.A. performed the fabrication and experiments. All authors were involved in conceiving the idea. All authors analyzed the data and contributed to writing the manuscript. A.H.S.-N and A.A. supervised the project. 

\section*{Competing interests}
The authors declare no competing interests.

\onecolumn
\newpage

%\centering

\begin{center}
  \section*{\textbf{\fontsize{16}{19.2}\selectfont Supplementary Information}}
\end{center} 

\bigskip \bigskip 
%\title{\textbf{\LARGE{Supplementary Information}}}
%\maketitle

\tableofcontents
\renewcommand\thefigure{S\arabic{figure}}
\setcounter{figure}{0}    
\section{Strain Profile in the Wafer}
\begin{figure*}[t!]
\centering
\includegraphics[width=0.8\textwidth]{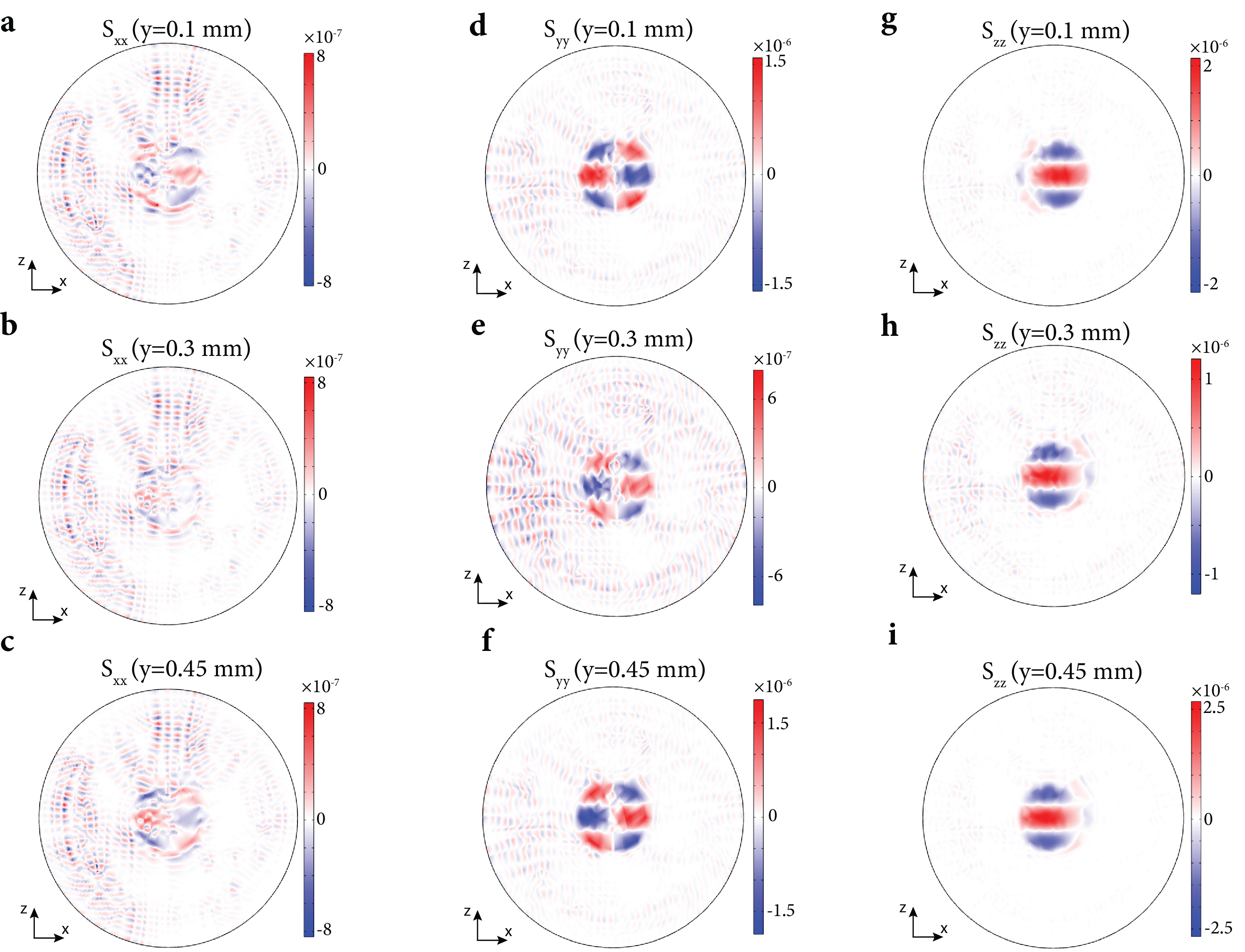}
\caption{\textbf{Normal strain profiles in the wafer}. \textbf{a}, $S_{xx}$ strain profile for the plane that is parallel and 0.1~mm above the bottom wafer surface. 2Vpp is applied to wafer surface electrodes. \textbf{b}, $S_{xx}$ strain profile for the plane that is parallel and 0.3~mm above the bottom wafer surface. 2Vpp is applied to wafer surface electrodes. \textbf{c}, $S_{xx}$ strain profile for the plane that is parallel and 0.45~mm above the bottom wafer surface. 2Vpp is applied to wafer surface electrodes. \textbf{d}, $S_{yy}$ strain profile for the plane that is parallel and 0.1~mm above the bottom wafer surface. 2Vpp is applied to wafer surface electrodes. \textbf{e}, $S_{yy}$ strain profile for the plane that is parallel and 0.3~mm above the bottom wafer surface. 2Vpp is applied to wafer surface electrodes. \textbf{f}, $S_{yy}$ strain profile for the plane that is parallel and 0.45~mm above the bottom wafer surface. 2Vpp is applied to wafer surface electrodes. \textbf{g}, $S_{zz}$ strain profile for the plane that is parallel and 0.1~mm above the bottom wafer surface. 2Vpp is applied to wafer surface electrodes. \textbf{h}, $S_{zz}$ strain profile for the plane that is parallel and 0.3~mm above the bottom wafer surface. 2Vpp is applied to wafer surface electrodes. \textbf{i}, $S_{zz}$ strain profile for the plane that is parallel and 0.45~mm above the bottom wafer surface. 2Vpp is applied to wafer surface electrodes.}
\label{fig:s1}
\end{figure*}

\begin{figure*}[t!]
\centering
\includegraphics[width=0.8\textwidth]{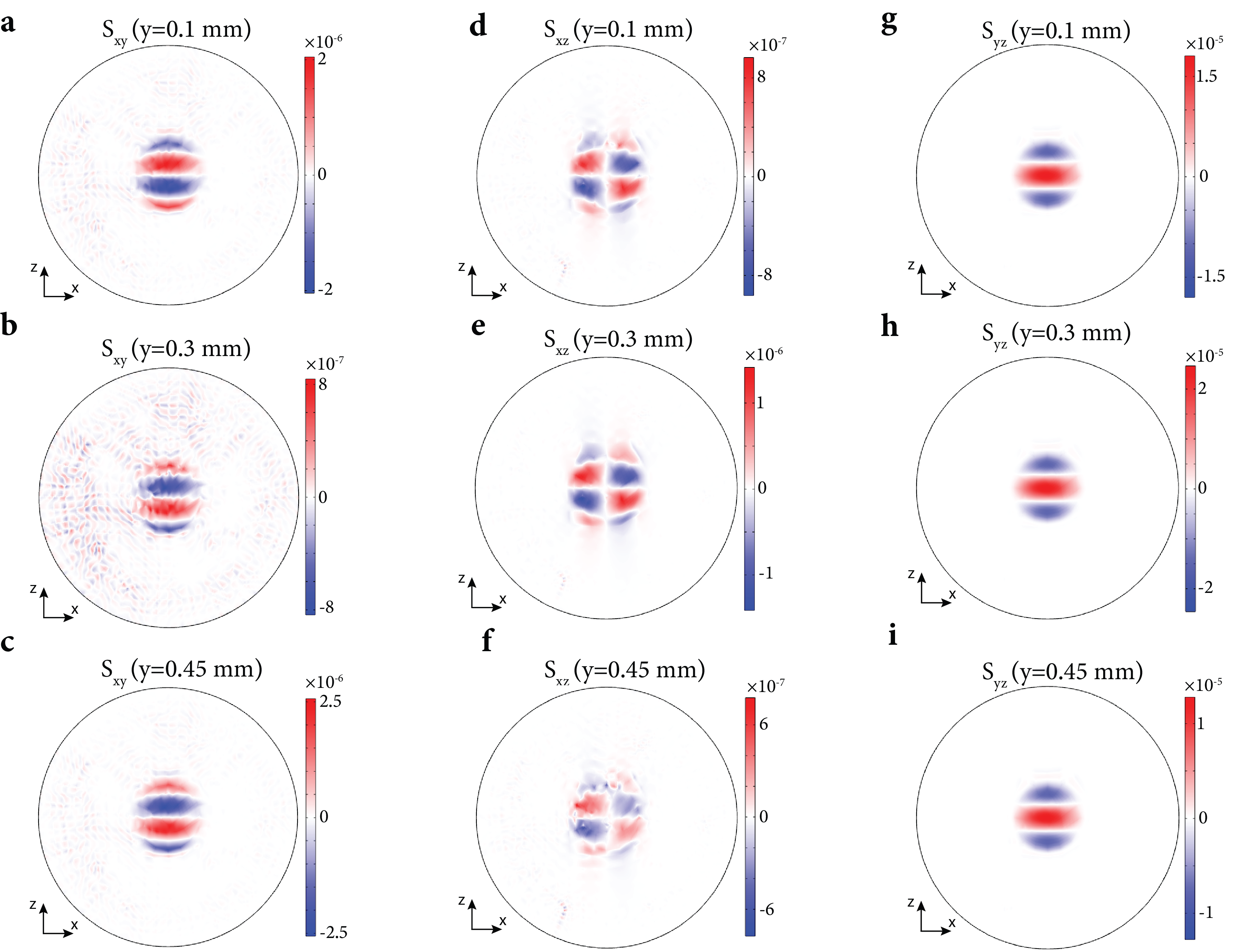}
\caption{\textbf{Shear strain profiles in the wafer}. \textbf{a}, $S_{xy}$ strain profile for the plane that is parallel and 0.1~mm above the bottom wafer surface. 2Vpp is applied to wafer surface electrodes. \textbf{b}, $S_{xy}$ strain profile for the plane that is parallel and 0.3~mm above the bottom wafer surface. 2Vpp is applied to wafer surface electrodes. \textbf{c}, $S_{xy}$ strain profile for the plane that is parallel and 0.45~mm above the bottom wafer surface. 2Vpp is applied to wafer surface electrodes. \textbf{d}, $S_{xz}$ strain profile for the plane that is parallel and 0.1~mm above the bottom wafer surface. 2Vpp is applied to wafer surface electrodes. \textbf{e}, $S_{xz}$ strain profile for the plane that is parallel and 0.3~mm above the bottom wafer surface. 2Vpp is applied to wafer surface electrodes. \textbf{f}, $S_{xz}$ strain profile for the plane that is parallel and 0.45~mm above the bottom wafer surface. 2Vpp is applied to wafer surface electrodes. \textbf{g}, $S_{yz}$ strain profile for the plane that is parallel and 0.1~mm above the bottom wafer surface. 2Vpp is applied to wafer surface electrodes. \textbf{h}, $S_{yz}$ strain profile for the plane that is parallel and 0.3~mm above the bottom wafer surface. 2Vpp is applied to wafer surface electrodes. \textbf{i}, $S_{yz}$ strain profile for the plane that is parallel and 0.45~mm above the bottom wafer surface. 2Vpp is applied to wafer surface electrodes.}
\label{fig:s2}
\end{figure*}

In this section, we show the strain profiles in the wafer when excited at the acoustic resonance frequency $f_c$ through the surface electrodes. We then derive the modified index ellipsoid due to strain. When the wafer is excited at $f_c = 3.7696~\text{MHz}$ with 2Vpp applied to the wafer electrodes, six different strain profiles are excited: $S_{xx}$, $S_{yy}$, $S_{zz}$, $S_{xy}$, $S_{xz}$, $S_{yz}$. These profiles are shown for three different planes at $y = 0.1~\text{mm}$, $y = 0.3~\text{mm}$, and $y = 0.45~\text{mm}$ in Fig.~\ref{fig:s1} and Fig.~\ref{fig:s2}, respectively. 

To calculate the overall rotation of polarization, the volume average strain for a localized region is used. This is because the wafer can be thought of as consisting of thin sheets of infinitesimal thickness parallel to the wafer surface. The volume average strain captures the overall rotation of polarization after the plane wave passes through these infinitesimal regions. The index ellipsoid for lithium niobate can be expressed as~\cite{crystal_optics}:

\begin{gather}
x^2\Big(\frac{1}{n_o^2} + p_{11}\bar{S}_{xx} + p_{12}\bar{S}_{yy}+ p_{13}\bar{S}_{zz}+2p_{14}\bar{S}_{yz}\Big) + y^2\Big(\frac{1}{n_o^2}+p_{12}\bar{S}_{xx}+p_{11}\bar{S}_{yy}+p_{13}\bar{S}_{zz}-2p_{14}\bar{S}_{yz}\Big) + \nonumber \\ z^2\Big(\frac{1}{n_e^2}+p_{13}\bar{S}_{xx}+p_{13}\bar{S}_{yy}+p_{33}\bar{S}_{zz}\Big) + 2yz\Big(p_{41}\bar{S}_{xx}-p_{41}\bar{S}_{yy}+2p_{44}\bar{S}_{yz}\Big) + \nonumber \\ 2zx\Big(2p_{44}\bar{S}_{xz}+2p_{41}\bar{S}_{xy}\Big) + 2xy\Big(2p_{14}\bar{S}_{xz}+(p_{11}-p_{12})\bar{S}_{xy}\Big) = 1 
\label{Eq.1} \tag{S1}
\end{gather}

To simplify the derivations, only the dominant strain distribution $S_{yz}$ contributing the most to the photoelastic interaction is used. Using the photoelastic coefficients from~\cite{photoelastic_coefficients}, this approximation simplifies the expression to:

\begin{gather}
x^2\Big(\frac{1}{n_o^2} +2p_{14}\bar{S}_{yz}\Big) + y^2\Big(\frac{1}{n_o^2} - 2p_{14}\bar{S}_{yz}\Big)  + z^2\Big(\frac{1}{n_e^2}\Big) + 2yz\Big(2p_{44}\bar{S}_{yz}\Big) = 1 
\label{Eq.2} \tag{S2}
\end{gather}

We apply a rotation to the yz axis such that the new form is diagonal~\cite{crystal_optics}. Using the coordinate transformations in equation \eqref{Eq.3}, equation \eqref{Eq.2} can be transformed into equation \eqref{Eq.4}.

\begin{gather}
y = y'\text{cos} \theta - z'\text{sin} \theta \nonumber \\ 
z = y'\text{sin} \theta + z'\text{cos} \theta \label{Eq.3} \tag{S3}
\end{gather}

\begin{gather}
x^2\Big(\frac{1}{n_o^2} +2p_{14}\bar{S}_{yz}\Big) + {y'}^2\Big(\frac{1}{n_o^2} -2p_{14}\bar{S}_{yz}+(2p_{44}\bar{S}_{yz})\text{tan} \theta\Big) + {z'}^2\Big(\frac{1}{n_e^2} - (2p_{44}\bar{S}_{yz})\text{tan} \theta\Big) = 1 \nonumber \\
\text{tan} (2 \theta) = \frac{4p_{44}\bar{S}_{yz}}{\Big(\frac{1}{n_o^2}-2p_{14}\bar{S}_{yz}\Big) - \Big(\frac{1}{n_e^2}\Big)} \label{Eq.4} \tag{S4}
\end{gather}

Since $\text{tan}\theta \ll 1$ in most cases ($\bar{S}_{yz} < 10^{-3}$ in experiments), we neglect the modulations of the $y'$ and $z'$ axis which include the $\text{tan}\theta$ term. The wafer can effectively be treated as three separate regions. The blue regions in Fig.~\ref{fig:s2}\textbf{g,h,i} are anti-phase with the red region. 

\begin{gather}
x^2\Big(\frac{1}{n_o^2} +2p_{14}\bar{S}_{yz}\Big) + {y'}^2\Big(\frac{1}{n_o^2}-2p_{14}\bar{S}_{yz}\Big) + {z'}^2\Big(\frac{1}{n_e^2}\Big) = 1 \label{Eq.5} \tag{S5}
\end{gather}

The wafer therefore functions as a polarization modulator at the applied frequency $f_c$. To simplify the derivations (and since $\text{tan}\theta << 1$), we assume $x' \approx x$ and $y' \approx y$ for the remainder of this document. 

\section{Intensity Modulation of a Laser Beam}
To convert the polarization modulator into an intensity modulator, the wafer is placed between polarizers. To find the response of an arbitrary laser beam passing through the polarizer, wafer, and polarizer, it is sufficient to find the intensity modulation profile for each plane wave (k-space), since a laser beam can be written as the linear superposition of plane waves. We will treat the propagation of a plane wave with wave vector ${\hat{k} = \hat{a}_x \text{sin}\theta \text{cos} \psi + \hat{a}_y \text{sin}\theta \text{sin} \psi + \hat{a}_z \text{cos}\theta}$, with random polarization, and intensity $I_0$ through the three elements separately. In each step, we will keep track of the polarization and intensity of the plane wave. The two polarizers have transmission axis $\hat{t} = \frac{\hat{a}_x + \hat{a}_z}{\sqrt{2}}$. The totally blocked polarization by the first polarizer is given by:  

\begin{gather}
\bar{p}_1 = \hat{t} \times \hat{k} = \hat{a}_x (-\text{sin}\theta \text{sin}\psi) + \hat{a}_y(\text{sin}\theta \text{cos}\psi - \text{cos}\theta) + \hat{a}_z(\text{sin}\theta \text{sin}\psi) \label{Eq.6} \tag{S6}
\end{gather}

The transmitted polarization is found as $\bar{p}_2 = \bar{p}_1 \times \hat{k}$. 

\begin{gather}
\bar{p}_2 = \hat{a}_x (\text{sin}\theta \text{cos}\theta \text{cos}\psi - \text{cos}^2 \theta - \text{sin}^2\theta \text{sin}^2\psi) + \hat{a}_y (\text{sin}^2\theta \text{cos}\psi \text{sin}\psi + \text{sin}\theta \text{cos}\theta \text{sin}\psi) + \hat{a}_z (\text{sin}\theta \text{cos}\theta \text{cos}\psi - \text{sin}^2\theta) \label{Eq.7} \tag{S7}
\end{gather}

Assuming random polarization for the incoming plane wave, the plane wave has intensity $\frac{I_0}{2}$ and polarization $\bar{p}_2$ after propagating through the first polarizer. Now we turn our attention to the propagation of the plane wave through the wafer. To simplify the calculations, we will first solve for the static case (no voltage applied to the wafer electrodes, and therefore no strain in the wafer). 

When a plane wave has oblique incidence on a uniaxial planar medium such as lithium niobate, an ordinary and an extraordinary wave is excited. To find the refraction angles and the effective refractive indices for these two possible solutions, we will use phase-matching. Let $\phi$ denote the angle between the surface normal of the wafer ($\hat{a}_y$) and $\hat{k}$. We can therefore write $\text{cos}\phi = \hat{k} \cdot \hat{a}_y = \text{sin}\theta \text{sin}\psi$, from which we find: $\phi = \text{cos}^{-1}(\text{sin}\theta \text{sin}\psi)$. Applying phase-matching for the ordinary and extraordinary waves:

\begin{gather}
\text{sin}\phi = n_o \text{sin}\theta_o = n_e(\theta_e)\text{sin} \tilde{\theta}_e \label{Eq.8} \tag{S8}
\end{gather}

In equation \eqref{Eq.8}, $\theta_e$ is the angle between $\hat{a}_z$ and the refracted extraordinary wave vector ($\hat{k}_{re}$), $\tilde{\theta}_e$ is the angle between $\hat{a}_y$ and $\hat{k}_{re}$. The extraordinary refractive index is expressed as: $\frac{1}{n^2(\theta_e)} = \frac{\text{cos}^2\theta_e}{n_o^2} + \frac{\text{sin}^2\theta_e}{n_e^2}$. 

The plane of incidence for the refraction problem is: $\bar{v} = \hat{a}_y \times \hat{k} = -\hat{a}_z \text{sin}\theta \text{cos}\psi + \hat{a}_x \text{cos}\theta$. We know that the refracted waves will lie on the plane of incidence. Therefore, $\hat{k}_{re} \cdot \bar{v} = \hat{k}_{ro} \cdot \bar{v} = 0$, and by definition of the angles, the following expressions are true: $\hat{k}_{re} \cdot \hat{a}_z = \text{cos}\theta_e$ and $\hat{k}_{re} \cdot \hat{a}_y = \text{cos}\tilde{\theta}_e$. We first solve for the extraordinary wave vector $\hat{k}_{re} = a_1 \hat{a}_x + a_2 \hat{a}_y + a_3 \hat{a}_z$. We use the following expressions to find $\hat{k}_{re}$:

\begin{gather}
\hat{k}_{re} \cdot \bar{v} = a_1 \text{cos}\theta -a_3 \text{sin}\theta \text{cos}\psi = 0 \nonumber \\ 
\hat{k}_{re} \cdot \hat{a}_z = a_3 = \text{cos}\theta_e \nonumber \\
\hat{k}_{re} \cdot \hat{a}_y = a_2 = \text{cos}\tilde{\theta}_e \nonumber \\
|\hat{k}_{re}| = \sqrt{a_1^2 + a_2^2 + a_3^2} = 1 \label{Eq.9} \tag{S9}
\end{gather}

We can use the above expressions to find:

\begin{gather}
a_1 = a_3 \text{tan}\theta \text{cos}\psi = \text{cos}\theta_e \text{tan}\theta \text{cos}\psi \nonumber \\
\text{cos}^2\theta_e \text{tan}^2\theta \text{cos}^2\psi + \text{cos}^2\tilde{\theta}_e + \text{cos}^2\theta_e = 1 \label{Eq.10} \tag{S10}
\end{gather}

Using the expression above, we can arrive at: $\text{sin}\tilde{\theta}_e = \text{cos}\theta_e \sqrt{\text{tan}^2\theta \text{cos}^2\psi + 1}$. Inserting this expression into the extraordinary phase-matching condition: $\text{sin}\phi = \frac{\text{cos}\theta_e\sqrt{\text{tan}^2\theta \text{cos}^2\psi + 1}}{\sqrt{\frac{\text{cos}^2\theta_e}{n_o^2} + \frac{\text{sin}^2\theta_e}{n_e^2}}}$.

To solve this, let $x = \text{cos}\theta_e$. The phase-matching condition can now be expressed as: ${\frac{\text{sin}\phi}{\sqrt{\text{tan}^2\theta \text{cos}^2\psi + 1}} = \frac{x}{\sqrt{\frac{x^2}{n_o^2} + \frac{1 - x^2}{n_e^2}}}}$. We therefore find the two angles as:

\begin{gather}
\theta_e = \text{cos}^{-1}\Bigg(\frac{n_o \text{sin}\phi}{\sqrt{n_o^2 n_e^2 (\text{tan}^2 \theta \text{cos}^2\psi + 1) - n_e^2 \text{sin}^2\phi + n_o^2 \text{sin}^2\phi}}\Bigg) \label{Eq.11} \tag{S11}
\end{gather}

\begin{gather}
\tilde{\theta}_e = \text{sin}^{-1}\Bigg(\frac{n_o \text{sin}\phi \sqrt{\text{tan}^2\theta \text{cos}^2\psi + 1}}{\sqrt{n_o^2 n_e^2 (\text{tan}^2 \theta \text{cos}^2\psi + 1) - n_e^2 \text{sin}^2\phi + n_o^2 \text{sin}^2\phi}}\Bigg) \label{Eq.12} \tag{S12}
\end{gather}

We now solve for the ordinary wave vector $\hat{k}_{ro} = b_1 \hat{a}_x + b_2 \hat{a}_y + b_3 \hat{a}_z$. We apply the same strategy we used to solve the extraordinary wave case to solve for the ordinary wave. We use the following to find $\hat{k}_{ro}$:

\begin{gather}
\hat{k}_{ro} \cdot \bar{v} = b_1 \text{cos}\theta - b_3 \text{sin}\theta \text{cos}\psi = 0 \nonumber \\
\hat{k}_{ro} \cdot \hat{a}_z = \text{cos}\theta_o \nonumber \\
\hat{k}_{ro} \cdot \hat{a}_y = \text{cos}\tilde{\theta}_o \nonumber \\
|\hat{k}_{ro}| = \sqrt{b_1^2 + b_2^2 + b_3^2} = 1 \label{Eq.13} \tag{S13}
\end{gather}

Using the expressions above, we can write: 

\begin{gather}
b_1 = b_3 \text{tan}\theta \text{cos}\psi = \text{cos}\theta_o \text{tan}\theta \text{cos}\psi \nonumber \\ \text{cos}^2\theta_o \text{tan}^2\theta \text{cos}^2\psi + \text{cos}^2\tilde{\theta}_o + \text{cos}^2\theta_o = 1
\label{Eq.14} \tag{S14}
\end{gather}

We can now arrive at $\text{sin}\tilde{\theta}_o = \text{cos}\theta_o \sqrt{\text{tan}^2\theta \text{cos}^2\psi + 1}$. Inserting this expression into the ordinary phase-matching condition: $\text{sin}\phi = n_o \text{cos}\theta_o \sqrt{\text{tan}^2\theta \text{cos}^2\psi + 1}$. We find the two angles as:

\begin{gather}
\theta_o = \text{cos}^{-1}\Bigg(\frac{\text{sin}\phi}{n_o \sqrt{\text{tan}^2\theta \text{cos}^2\psi + 1}}\Bigg) \label{Eq.15} \tag{S15}
\end{gather}

\begin{gather}
\tilde{\theta}_o = \text{sin}^{-1}\Big(\frac{\text{sin}\phi}{n_o}\Big) \label{Eq.16} \tag{S16}
\end{gather}

We will now find the polarizations $\bar{p}_o$ and $\bar{p}_e$ corresponding to the ordinary and extraordinary refracted waves, respectively. 

\begin{gather}
\bar{p}_o = \hat{k}_{ro} \times \hat{a}_z = \text{cos}\Bigg(\text{sin}^{-1}\Big(\frac{\text{sin}\phi}{n_o}\Big)\Bigg)\hat{a}_x  -\frac{\text{sin}\phi \text{tan}\theta \text{cos}\psi}{n_o \sqrt{\text{tan}^2\theta \text{cos}^2\psi + 1}}\hat{a}_y   \label{Eq.17} \tag{S17}
\end{gather}

\begin{gather}
\bar{p}_e = (\hat{k}_{re} \times \hat{a}_z) \times \hat{k}_{re} = -\text{cos}^2\theta_e \text{tan}\theta \text{cos}\psi \hat{a}_x - \text{cos}\theta_e \text{cos}\tilde{\theta}_e \hat{a}_y + \Big(\text{cos}^2\tilde{\theta}_e + \text{cos}^2\theta_e \text{tan}^2\theta \text{cos}^2\psi\Big)\hat{a}_z \label{Eq.18} \tag{S18}
\end{gather}

We will now focus on finding the polarizations $\bar{p}_{oi}$ and $\bar{p}_{ei}$ in air incident to the wafer surface that correspond to the ordinary and extraordinary polarizations in the wafer: $\bar{p}_o$ and $\bar{p}_e$. We assume for simplifying the derivations that there is no reflection at the lithium niobate-air interface (ideal anti-reflection coating assumption). This assumption is fairly accurate if the incidence angle is nearly perpendicular to the wafer surface. This assumption allows us to simplify the refraction problem; polarization $\bar{p}_{oi}$ gets mapped to $\bar{p}_{o}$ in the wafer and polarization $\bar{p}_{ei}$ gets mapped to $\bar{p}_{e}$ in the wafer. We will first solve for the ordinary wave. Using the assumption of no reflection at the air-wafer boundary, we use the following to find $\bar{p}_{oi}$:

\begin{gather}
\bar{p}_{oi} \cdot \hat{k} = 0 \nonumber \\
\bar{p}_o \cdot \bar{v} = \bar{p}_{oi} \cdot \bar{v} = \frac{\text{cos}\theta \text{cos}\Big(\text{sin}^{-1}\big(\frac{\text{sin}\phi}{n_o}\big)\Big)}{\sqrt{\text{cos}^2\Big(\text{sin}^{-1}\big(\frac{\text{sin}\phi}{n_o}\big)\Big) + \frac{\text{sin}^2\phi \text{tan}^2\theta \text{cos}^2\psi}{n_o^2(\text{tan}^2\theta \text{cos}^2\psi + 1)}}} \label{Eq.19} \tag{S19}
\end{gather}

Let $\hat{p}_{oi} = c_1 \hat{a}_x + c_2\hat{a}_y + c_3 \hat{a}_z$, where $c_1^2 + c_2^2 + c_3^2 = 1$. We can now arrive at:

\begin{gather}
c_1 \text{sin}\theta \text{cos}\psi + c_2 \text{sin}\theta \text{sin}\psi + c_3 \text{cos}\theta = 0 \nonumber \\
c_1\text{cos}\theta - c_3\text{sin}\theta \text{cos}\psi = \frac{\text{cos}\theta \text{cos}\Big(\text{sin}^{-1}\big(\frac{\text{sin}\phi}{n_o}\big)\Big)}{\sqrt{\text{cos}^2\Big(\text{sin}^{-1}\big(\frac{\text{sin}\phi}{n_o}\big)\Big) + \frac{\text{sin}^2\phi \text{tan}^2\theta \text{cos}^2\psi}{n_o^2(\text{tan}^2\theta \text{cos}^2\psi + 1)}}} \label{Eq.20} \tag{S20}
\end{gather}

Using the expressions above, we find $c_2$:
\begin{gather}
c_2 = \frac{-c_3(\text{cos}\theta + \text{sin}\theta \text{tan}\theta \text{cos}^2\psi)}{\text{sin}\theta \text{sin}\psi} - \frac{\text{cot}\psi \text{cos}\Big(\text{sin}^{-1}\big(\frac{\text{sin}\phi}{n_o}\big)\Big)}{\sqrt{\text{cos}^2\Big(\text{sin}^{-1}\big(\frac{\text{sin}\phi}{n_o}\big)\Big) + \frac{\text{sin}^2\phi \text{tan}^2\theta \text{cos}^2\psi}{n_o^2(\text{tan}^2\theta \text{cos}^2\psi + 1)}}} \label{Eq.21} \tag{S21}
\end{gather}

Using $c_1^2 + c_2^2 + c_3^2 = 1$, $c_3$ can be expressed as the solution to the following quadratic equation: $\tilde{a}_1c_3^2 + \tilde{a}_2c_3 + \tilde{a}_3 = 0$, where: 

\begin{gather}
\tilde{a}_1 = \text{tan}^2\theta \text{cos}^2\psi + \frac{(\text{cos}\theta + \text{sin}\theta \text{tan}\theta \text{cos}^2\psi)^2}{\text{sin}^2\theta \text{sin}^2\psi} + 1 \nonumber \\ \tilde{a}_2 = \frac{2\text{cos}\Big(\text{sin}^{-1}\big(\frac{\text{sin}\phi}{n_o}\big)\Big)}{\sqrt{\text{cos}^2\Big(\text{sin}^{-1}\big(\frac{\text{sin}\phi}{n_o}\big)\Big) + \frac{\text{sin}^2\phi \text{tan}^2\theta \text{cos}^2\psi}{n_o^2(\text{tan}^2\theta \text{cos}^2\psi + 1)}}}\Bigg(\text{tan}\theta \text{cos}\psi + \frac{\text{cot}\psi (\text{cos}\theta + \text{sin}\theta \text{tan}\theta \text{cos}^2\psi)}{\text{sin}\theta \text{sin}\psi}\Bigg) \nonumber \\
\tilde{a}_3 = \frac{\text{cos}^2\Big(\text{sin}^{-1}\big(\frac{\text{sin}\phi}{n_o}\big)\Big) + \text{cot}^2\psi \text{cos}^2\Big(\text{sin}^{-1}\big(\frac{\text{sin}\phi}{n_o}\big)\Big)}{\text{cos}^2(\text{sin}^{-1}(\frac{\text{sin}\phi}{n_o})) + \frac{\text{sin}^2\phi \text{tan}^2\theta \text{cos}^2\psi}{n_o^2(\text{tan}^2\theta \text{cos}^2\psi + 1)}} - 1 \label{Eq.22} \tag{S22}
\end{gather}

The solution to this quadratic equation is: $c_3 = \frac{-\tilde{a}_2 \pm \sqrt{\tilde{a}_2^2 - 4\tilde{a}_1 \tilde{a}_3}}{2\tilde{a}_1}$ Since the quadratic equation can have two real solutions, how do we pick the right one? The solution that we pick is the one that maximizes $\hat{p}_{oi} \cdot \bar{p}_o$. The reason why two solutions emerge is because of the way we have framed the question. Specifically, using the equality $\hat{p}_o \cdot \bar{v} = \hat{p}_{oi} \cdot \bar{v}$, there are two angles satisfying the dot product equality. What we were trying to enforce was that the angle between $\hat{p}_o$ and $\bar{v}$ is equal to the angle between $\hat{p}_{oi}$ and $\bar{v}$, since no reflection at the boundary was assumed. We can now find $c_2$ by plugging $c_3$ into equation~\eqref{Eq.21}. $c_1$ is found as:

\begin{gather}
c_1 = \sqrt{1 - c_2^2 - c_3^2} \label{Eq.23} \tag{S23}
\end{gather}

We will now find $\hat{p}_{ei}$. Let $\hat{p}_{ei} = d_1\hat{a}_x + d_2\hat{a}_y + d_3\hat{a}_z$, where $d_1^2 + d_2^2 + d_3^2 = 1$ and $\bar{p}_{ei} = \hat{p}_{oi} \times \hat{k}$. We can find the components as follows:

\begin{gather}
d_1 = \frac{c_2 \text{cos}\theta - c_3 \text{sin}\theta \text{sin}\psi}{\sqrt{(c_2 \text{cos}\theta -c_3 \text{sin}\theta \text{sin}\psi)^2 + (-c_1 \text{cos}\theta + c_3 \text{sin}\theta \text{cos}\psi)^2 + (c_1 \text{sin}\theta \text{sin}\psi - c_2 \text{sin}\theta \text{cos}\psi)^2}} \label{Eq.24} \tag{S24}
\end{gather}

\begin{gather}
d_2 = \frac{-c_1 \text{cos}\theta + c_3 \text{sin}\theta \text{cos}\psi}{\sqrt{(c_2 \text{cos}\theta -c_3 \text{sin}\theta \text{sin}\psi)^2 + (-c_1 \text{cos}\theta + c_3 \text{sin}\theta \text{cos}\psi)^2 + (c_1 \text{sin}\theta \text{sin}\psi - c_2 \text{sin}\theta \text{cos}\psi)^2}} \label{Eq.25} \tag{S25}
\end{gather}

\begin{gather}
d_3 = \frac{c_1 \text{sin}\theta \text{sin}\psi - c_2 \text{sin}\theta \text{cos}\psi}{\sqrt{(c_2 \text{cos}\theta -c_3 \text{sin}\theta \text{sin}\psi)^2 + (-c_1 \text{cos}\theta + c_3 \text{sin}\theta \text{cos}\psi)^2 + (c_1 \text{sin}\theta \text{sin}\psi - c_2 \text{sin}\theta \text{cos}\psi)^2}} \label{Eq.26} \tag{S26}
\end{gather}

We can express $\hat{p}_2$ as a linear superposition of $\bar{p}_{oi}$ and $\bar{p}_{ei}$. Specifically, $\hat{p}_2 = c_o\hat{p}_{oi} + c_e\hat{p}_{ei}$, where $c_o$ and $c_e$ denote the ordinary and extraordinary wave amplitudes in air, respectively. We therefore find $c_o = \hat{p}_2 \cdot \hat{p}_{oi}$ and $c_e = \hat{p}_2 \cdot \hat{p}_{ei}$. 

\begin{gather}
c_o = \frac{c_1 (\text{sin}\theta \text{cos}\theta \text{cos}\psi - \text{cos}^2\theta - \text{sin}^2\theta \text{sin}^2\psi) + c_2(\text{sin}^2\theta \text{cos}\psi \text{sin}\psi + \text{sin}\theta \text{cos}\theta \text{sin}\psi) + c_3(\text{sin}\theta \text{cos}\theta \text{cos}\psi - \text{sin}^2\theta)}{\sqrt{(\text{sin}\theta \text{cos}\theta \text{cos}\psi - \text{cos}^2\theta - \text{sin}^2\theta \text{sin}^2\psi)^2 + (\text{sin}^2\theta \text{cos}\psi \text{sin}\psi + \text{sin}\theta \text{cos}\theta \text{sin}\psi)^2 + (\text{sin}\theta \text{cos}\theta \text{cos}\psi - \text{sin}^2\theta)^2}} \label{Eq.27} \tag{S27}
\end{gather}

\begin{gather}
c_e = \frac{d_1 (\text{sin}\theta \text{cos}\theta \text{cos}\psi - \text{cos}^2\theta - \text{sin}^2\theta \text{sin}^2\psi) + d_2(\text{sin}^2\theta \text{cos}\psi \text{sin}\psi + \text{sin}\theta \text{cos}\theta \text{sin}\psi) + d_3(\text{sin}\theta \text{cos}\theta \text{cos}\psi - \text{sin}^2\theta)}{\sqrt{(\text{sin}\theta \text{cos}\theta \text{cos}\psi - \text{cos}^2\theta - \text{sin}^2\theta \text{sin}^2\psi)^2 + (\text{sin}^2\theta \text{cos}\psi \text{sin}\psi + \text{sin}\theta \text{cos}\theta \text{sin}\psi)^2 + (\text{sin}\theta \text{cos}\theta \text{cos}\psi - \text{sin}^2\theta)^2}} \label{Eq.28} \tag{S28}
\end{gather}

Now we will calculate the refractive indices experienced by the ordinary and extraordinary waves when the wafer is excited through its surface electrodes with frequency $f_c$, leading to volume average strain $\bar{S}_{yz}$ in the wafer. To simplify the expressions, we assume that the polarization directions $\bar{p_o}$ and $\bar{p}_e$ do not change when strain is present in the wafer. We will only calculate the change in the refractive indices using this assumption (due to change in the index ellipsoid via photoelasticity) to simplify the derivation. This assumption does not change the results significantly, since the dominant contribution is from the refractive index change.

We will first solve for the ordinary wave. We need to find the intersection of the ellipsoid $x^2\Big(\frac{1}{n_o^2} + 2p_{14}\bar{S}_{yz}\Big) + y^2\Big(\frac{1}{n_o^2} - 2p_{14}\bar{S}_{yz}\Big) = 1$ and the vector $a\Big(\text{cos}\Big(\text{sin}^{-1}\big(\frac{\text{sin}\phi}{n_o}\big)\Big)\hat{a}_x -\frac{\text{sin}\phi \text{tan}\theta \text{cos}\psi}{n_o\sqrt{\text{tan}^2\theta \text{cos}^2\psi + 1}}\hat{a}_y\Big)$. This vector is a scaled form of $\bar{p}_{o}$ with the scalar $a$, and the length of this vector is equal to the refractive index $n_o(t)$ experienced by the ordinary wave. Using the intersection of the ellipsoid and the vector, we arrive at: $a^2\text{cos}^2\Big(\text{sin}^{-1}\big(\frac{\text{sin}\phi}{n_o}\big)\Big)\Big(\frac{1}{n_o^2} + 2p_{14}\bar{S}_{yz}\Big) + \frac{a^2\text{sin}^2\phi \text{tan}^2\theta \text{cos}^2\psi}{n_o^2(\text{tan}^2\theta \text{cos}^2\psi + 1)}\Big(\frac{1}{n_o^2} - 2p_{14}\bar{S}_{yz}\Big) = 1$. The length of the scaled vector can be found as:

\begin{gather}
n_o^2(t) = \frac{n_o^2}{1 + \frac{2p_{14}\bar{S}_{yz}n_o^2(n_o^2\text{cos}^2(\text{sin}^{-1}(\frac{\text{sin}\phi}{n_o}))(\text{tan}^2\theta \text{cos}^2\psi + 1)) - \text{sin}^2\phi \text{tan}^2\theta \text{cos}^2\psi}{n_o^2(\text{tan}^2\theta \text{cos}^2\psi + 1)\text{cos}^2(\text{sin}^{-1}(\frac{\text{sin}\phi}{n_o})) + \text{sin}^2\phi \text{tan}^2\theta \text{cos}^2\psi}} \label{Eq.29} \tag{S29}
\end{gather}

\begin{gather}
n_o(t) \approx n_o\Bigg(1 - \frac{p_{14}\bar{S}_{yz}n_o^2\Big(n_o^2\text{cos}^2(\text{sin}^{-1}(\frac{\text{sin}\phi}{n_o}))(\text{tan}^2\theta \text{cos}^2\psi + 1) - \text{sin}^2\phi \text{tan}^2\theta \text{cos}^2\psi\Big)}{n_o^2(\text{tan}^2\theta \text{cos}^2\psi + 1)\text{cos}^2(\text{sin}^{-1}(\frac{\text{sin}\phi}{n_o})) + \text{sin}^2\phi \text{tan}^2\theta \text{cos}^2\psi}\Bigg) \label{Eq.30} \tag{S30}
\end{gather}

We now solve for the extraordinary wave. We need to find the intersection of the ellipsoid $x^2\Big(\frac{1}{n_o^2} + 2p_{14}\bar{S}_{yz}\Big) + y^2\Big(\frac{1}{n_o^2} - 2p_{14}\bar{S}_{yz}\Big) + z^2\Big(\frac{1}{n_e^2}\Big) = 1$ and the vector $b\big(-\text{cos}^2\theta_e \text{tan}\theta \text{cos}\psi\hat{a}_x -\text{cos}\tilde{\theta}_e \text{cos}\theta_e\hat{a}_y + (\text{cos}^2\tilde{\theta}_e + \text{cos}^2\theta_e \text{tan}^2\theta \text{cos}^2\psi)\hat{a}_z\big)$. This vector is a scaled form of $\bar{p}_e$ with the scalar $b$. Using the intersection of the ellipsoid and the vector, we arrive at: $b^2(\text{cos}^4\theta_e \text{tan}^2\theta \text{cos}^2\psi)(\frac{1}{n_o^2} + 2p_{14}\bar{S}_{yz}) + b^2 \text{cos}^2\tilde{\theta}_e(\frac{1}{n_o^2} - 2p_{14}\bar{S}_{yz}) + b^2(\text{cos}^2\tilde{\theta}_e + \text{cos}^2\theta_e \text{tan}^2\theta \text{cos}^2\psi)^2(\frac{1}{n_e^2}) = 1$. The length of the scaled vector can be found as: 

\begin{gather}
n_e^2(t) = \frac{n_o^2 n_e^2}{n_o^2 \text{sin}^2\theta_e + n_e^2 \text{cos}^2\theta_e + \frac{2p_{14}\bar{S}_{yz}n_o^2(n_e^2 \text{cos}^4\theta_e \text{tan}^2\theta \text{cos}^2\psi - \text{cos}^2\tilde{\theta_e} \text{cos}^2\theta_e n_e^2)}{\text{cos}^4\theta_e \text{tan}^2\theta \text{cos}^2\psi + \text{cos}^2\tilde{\theta}_e \text{cos}^2\theta_e + (\text{cos}^2\tilde{\theta_e} + \text{cos}^2\theta_e \text{tan}^2\theta \text{cos}^2\psi)^2}} \label{Eq.31} \tag{S31}
\end{gather}

\begin{gather}
n_e(t) \approx \frac{n_o n_e}{\sqrt{n_o^2 \text{sin}^2\theta_e + n_e^2 \text{cos}^2\theta_e}}\Bigg(1 - \frac{p_{14}\bar{S}_{yz}n_o^2\big(n_e^2 \text{cos}^4\theta_e \text{tan}^2\theta \text{cos}^2\psi - \text{cos}^2\tilde{\theta}_e \text{cos}^2\theta_e n_e^2\big)}{(n_o^2 \text{sin}^2\theta_e + n_e^2 \text{cos}^2\theta_e)(\text{cos}^4\theta_e \text{tan}^2\theta \text{cos}^2\psi + \text{cos}^2\tilde{\theta}_e \text{cos}^2\theta_e + (\text{cos}^2\tilde{\theta}_e + \text{cos}^2\theta_e \text{tan}^2\theta \text{cos}^2\psi)^2)}\Bigg) \label{Eq.32} \tag{S32}
\end{gather}

We can now calculate the time-varying phase picked up by the ordinary and extraordinary waves after propagating through the wafer as $\phi_o(t)$ and $\phi_e(t)$, respectively. These are expressed as follows:

\begin{gather}
\phi_o(t) = \frac{2 \pi L}{\lambda}\text{cos}(\tilde{\theta}_o)n_o(t) \label{Eq.33} \tag{S33}
\end{gather}

\begin{gather}
\phi_e(t) = \frac{2 \pi L}{\lambda}\text{cos}(\tilde{\theta}_e)n_e(t) \label{Eq.34} \tag{S34}
\end{gather}

For the expressions above, $\lambda$ is the free-space wavelength of the plane wave and $L$ is the thickness of the wafer (parallel to $\hat{a}_y$). We can express the static ($\phi_s$) and dynamic ($\phi_D$) phase difference as: 

\begin{gather}
\phi_s + \phi_D \text{cos}(2 \pi f_c t) = \phi_o(t) - \phi_e(t) \label{Eq.35} \tag{S35}
\end{gather}

We will now turn our attention to the second component, the second polarizer (analyzer). The electric field incident on the second polarizer is given as: 

\begin{gather}
\bar{E}_1(t) = \sqrt{\frac{I_0}{2}}\Big(\hat{p}_{oi}c_oe^{jw_Lt + \phi_o(t)} + \hat{p}_{ei}c_e^{jw_Lt + \phi_e(t)}\Big) \label{Eq.36} \tag{S36}
\end{gather}

\begin{figure*}[t!]
\centering
\includegraphics[width=0.8\textwidth]{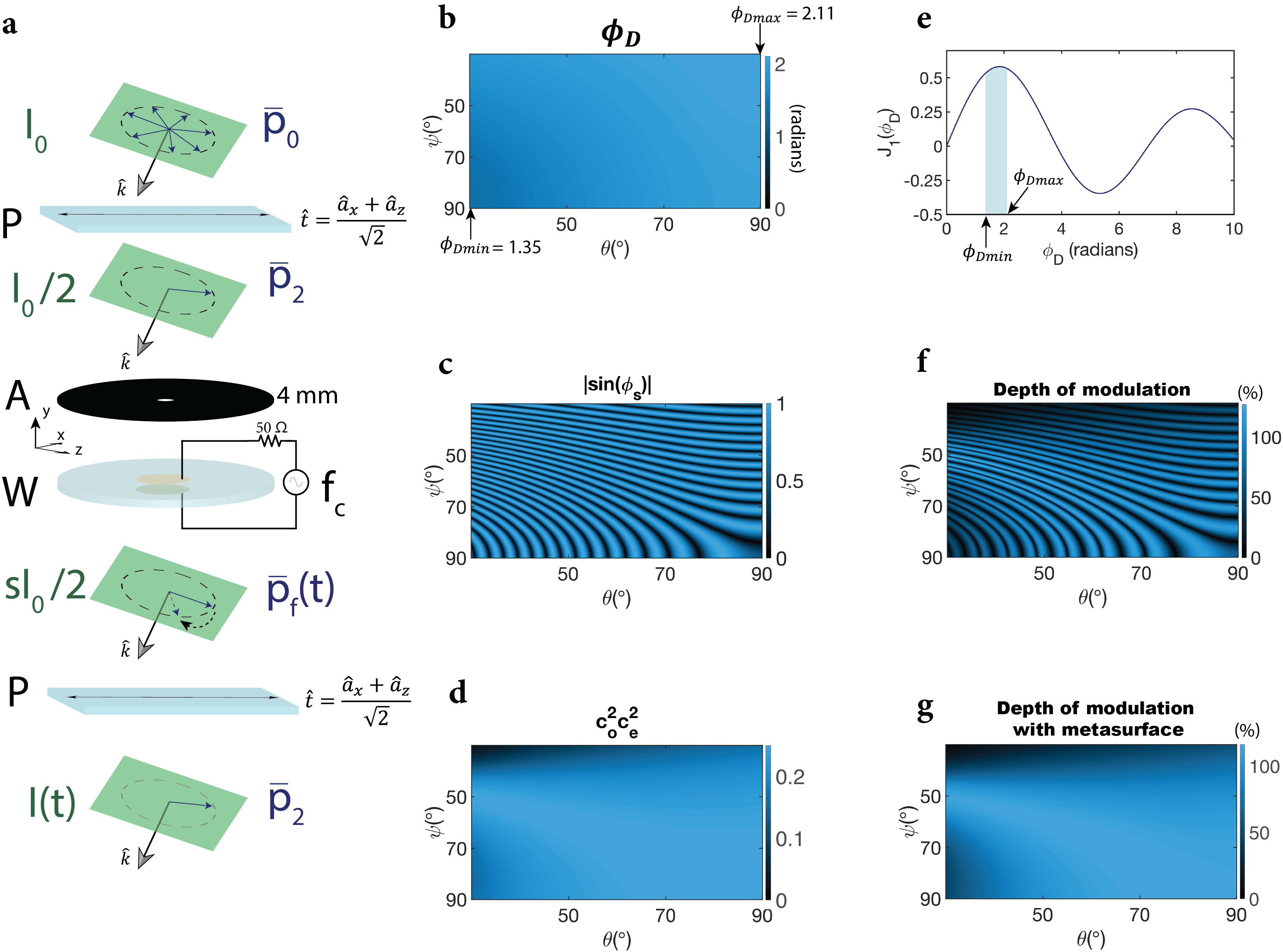}
\caption{\textbf{Intensity modulation of a plane wave for optimal strain in the wafer}. \textbf{a}, The intensity modulation for a plane wave that is incident on the intensity modulator with wave vector $\hat{k}$ is shown. Standard spherical coordinate notation is used for the angles to define $\hat{k} = \hat{a}_x \text{sin}\theta \text{cos}\psi + \hat{a}_y \text{sin}\theta \text{sin}\psi + \hat{a}_z \text{cos}\theta$. The plane wave has random polarization $\bar{p}_0$ and intensity $I_0$. After passing through the first polarizer (P) with transmission axis $\hat{t} = (\hat{a}_x + \hat{a}_z)/\sqrt{2}$, the intensity of the plane wave is reduced to $I_0/2$, with a polarization direction of $\bar{p}_2$. The plane wave then passes through an aperture (A) with an aperture diameter of 4~mm, so that only the center part of the $S_{yz}$ strain profile is used. The plane wave then passes through the wafer (W) that has a volume average strain $\bar{S}_{yz} = 5.67 \times 10^{-4}$ at $f_c = 3.7696~\text{MHz}$. The plane wave that has propagated through the wafer has intensity reduced to $s I_0/2$, with time-dependent polarization $\bar{p}_f(t)$. $0< s < 1$ captures the attenuation of the plane wave due to passing through the aperture. The polarization rotation of the plane wave is converted into intensity modulation after passing through the second polarizer (P). The intensity of the plane wave $I(t)$ is now time-dependent and has a polarization direction of $\bar{p}_2$. \textbf{b}, The dynamic phase accumulated ($\phi_D$) by plane waves incident at different angles to the wafer is shown. The minimum and maximum values attained by $\phi_D$ are $\phi_{Dmin} = 1.35$ and $\phi_{Dmax} = 2.11$, respectively. \textbf{c}, $|\text{sin}(\phi_s)|$ for plane waves incident at different angles to the wafer is shown, where $\phi_s$ is the static phase accumulated by the plane waves. \textbf{d}, $c_o^2 c_e^2$ for plane waves incident at different angles to the wafer is shown, where $c_o$ is the amplitude of the excited ordinary wave, and $c_e$ is the amplitude of the excited extraordinary wave. \textbf{e}, $J_1(\phi_D)$ is shown as a function of $\phi_D$. The maximum and minimum for $J_1(\phi_D)$ corresponding to the maximum and minimum $\phi_D$ in \textbf{b} is shown. \textbf{f}, Depth of modulation (DoM) as a percentage is shown for plane waves incident on the wafer at different angles. \textbf{g}, Depth of modulation (DoM) as a percentage is shown for plane waves incident on the wafer at different angles. An ideal polarization manipulating metasurface is assumed to be coated on the wafer surface such that $\text{sin}\phi_s = 1 \ \forall \ (\theta,\psi)$.}
\label{fig:s3}
\end{figure*}

\begin{figure*}[t!]
\centering
\includegraphics[width=0.8\textwidth]{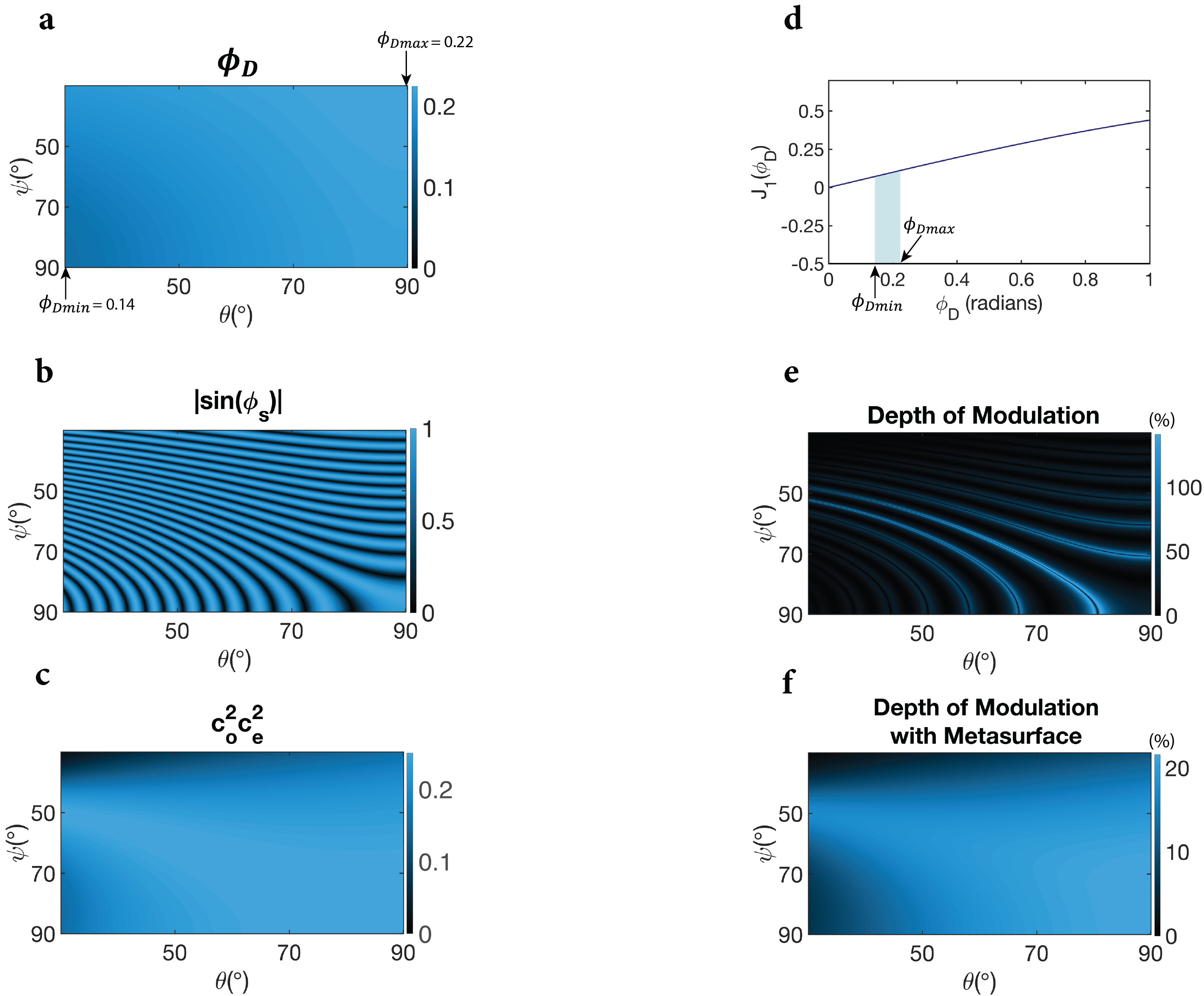}
\caption{\textbf{Intensity modulation of a plane wave. Volume average strain $\boldsymbol{\bar{S}_{yz} = 6 \times 10^{-5}}$ at $\boldsymbol{f_c = 3.7696~\text{MHz}}$ is used for this figure. This volume average strain is computed for a diameter of 4~mm centered on the wafer.} \textbf{a}, The dynamic phase accumulated ($\phi_D$) by plane waves incident at different angles to the wafer is shown. The minimum and maximum values attained by $\phi_D$ are $\phi_{Dmin} = 0.14$ and $\phi_{Dmax} = 0.22$, respectively. \textbf{b}, $|\text{sin}(\phi_s)|$ for plane waves incident at different angles to the wafer is shown, where $\phi_s$ is the static phase accumulated by the plane waves. \textbf{c}, $c_o^2 c_e^2$ for plane waves incident at different angles to the wafer is shown, where $c_o$ is the amplitude of the excited ordinary wave, and $c_e$ is the amplitude of the excited extraordinary wave. \textbf{d}, $J_1(\phi_D)$ is shown as a function of $\phi_D$. The maximum and minimum for $J_1(\phi_D)$ corresponding to the maximum and minimum $\phi_D$ in \textbf{a} is shown. \textbf{e}, Depth of modulation (DoM) as a percentage is shown for plane waves incident on the wafer at different angles. \textbf{f}, Depth of modulation (DoM) as a percentage is shown for plane waves incident on the wafer at different angles. An ideal polarization manipulating metasurface is assumed to be coated on the wafer surface such that $\text{sin}\phi_s = 1 \ \forall \ (\theta,\psi)$.}
\label{fig:s4}
\end{figure*}

\begin{figure*}[t!]
\centering
\includegraphics[width=0.8\textwidth]{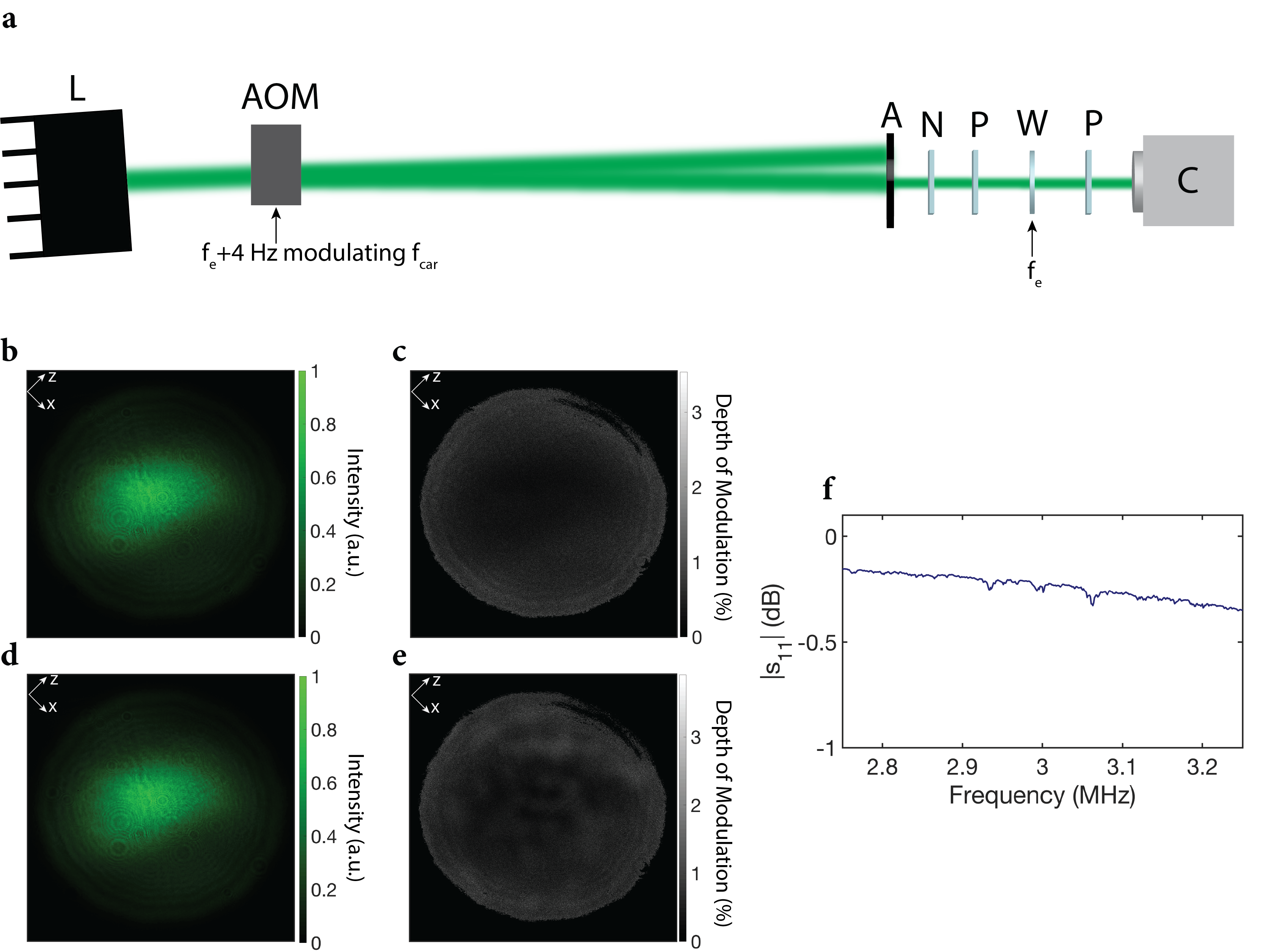}
\caption{\textbf{Checking contribution of electro-optic effect}. \textbf{a}, Schematic of the characterization setup is shown. The setup includes a laser (L) with a wavelength of 532~nm that passes through a free-space acousto-optic modulator (AOM). The laser beam is intensity modulated at $f_e + 4~\text{Hz} = 3.000004~\text{MHz}$ by modulating the carrier frequency $f_{car} = 80~\text{MHz}$ exciting the AOM. The setup also includes an aperture (A) with a diameter of 9.3~mm, a neutral density filter (N), two polarizers (P) with transmission axis $\hat{t}$, wafer (W), and a standard CMOS camera (C). The camera detects the intensity modulated laser beam with a frame rate of 30~Hz. \textbf{b}, Time-averaged intensity profile of the laser beam detected by the camera is shown for $\phi = 0$ and when no voltage is applied to wafer surface electrodes. \textbf{c}, The DoM at 4~Hz of the laser beam is shown per pixel for $\phi = 0$ and when no voltage is applied to wafer surface electrodes. \textbf{d}, Time-averaged intensity profile of the laser beam detected by the camera is shown for $\phi = 0$ and when 90~mW of RF power at $f_e = 3~\text{MHz}$ is applied to wafer surface electrodes. \textbf{e}, The DoM at 4~Hz of the laser beam is shown per pixel for $\phi = 0$ and when 90~mW of RF power at $f_e = 3~\text{MHz}$ is applied to wafer surface electrodes. \textbf{f}, $s_{11}$ scan with respect to $50~ \Omega$ and using 0~dBm excitation power with a bandwidth of 100~Hz is shown around 3~MHz.}
\label{fig:s5}
\end{figure*}

In the equation above, $w_L$ is the angular frequency of the optical field. After propagating through the second polarizer, the electric field becomes: 

\begin{gather}
\bar{E}_2(t) = \sqrt{\frac{I_0}{2}}\Big((\hat{p}_{oi} \cdot \hat{p}_{2}) \hat{p}_{2} c_oe^{jw_Lt + \phi_o(t)} + (\hat{p}_{ei} \cdot \hat{p}_{2}) \hat{p}_{2} c_ee^{jw_Lt + \phi_e(t)}\Big) = \sqrt{\frac{I_0}{2}}\Big(\hat{p}_{2} c_o^2e^{jw_Lt + \phi_o(t)} +  \hat{p}_{2} c_e^2e^{jw_Lt + \phi_e(t)}\Big) \label{Eq.37} \tag{S37}
\end{gather}

The intensity of the plane wave that has propagated through the second polarizer is given as follows, which is written in terms of the harmonics of $f_c$ using the Jacobi-Anger expansion:

\begin{gather}
I(t) = \Big|\bar{E}_2(t)\Big|^2 =  \frac{I_0}{2}\Big(c_o^4 + c_e^4 + 2c_o^2c_e^2 \text{cos}\big(\phi_s + \phi_D \text{cos}(2 \pi f_c t)\big)\Big) \nonumber \\
= \frac{I_0}{2}\Bigg(c_o^4 + c_e^4 + 2c_o^2c_e^2\bigg(\text{cos}\phi_s \Big(J_0(\phi_D) + 2  \sum_{n=1}^{\infty} (-1)^n J_{2n}(\phi_D) \text{cos}(4 \pi f_c t)\Big) + 2\text{sin}\phi_s \Big(\sum_{n=1}^{\infty} (-1)^n J_{2n-1}(\phi_D)\text{cos}((2n-1)2 \pi f_ct)\Big)\bigg)\Bigg)\label{Eq.38} \tag{S38}
\end{gather}

Now we will look into the intensity modulation term $P_{f_c}$ at frequency $f_c$, and the constant term (DC). These are expressed as follows:

\begin{gather}
\text{DC} = \frac{I_0}{2}\Big(c_o^4 + c_e^4 + 2c_o^2 c_e^2 J_0(\phi_D) \text{cos}\phi_s\Big) \nonumber \\
P_{f_c} = 2I_0c_o^2 c_e^2 J_1(\phi_D) \text{sin}\phi_s
\label{Eq.39} \tag{S39}
\end{gather}

We can define the depth of modulation (DoM) as: 

\begin{gather}
\text{DoM} = \frac{4c_o^2 c_e^2 J_1(\phi_D) \text{sin}\phi_s}{c_o^4 + c_e^4 + 2c_o^2 c_e^2 J_0(\phi_D) \text{cos}\phi_s} \label{Eq.40} \tag{S40}
\end{gather}

There are two important observations to make regarding equation \eqref{Eq.39} and equation \eqref{Eq.40}. First, $P_{f_c}$ can be greater than $\frac{I_0}{4}$ since $J_1(\phi_D)$ can exceed 0.5. This means that for the same peak-to-peak variation as a pure sinusoid at frequency $f_c$, the power in the fundamental of the Bessel function is larger. This is similar to how a square wave has a fundamental tone larger than a sinusoid with the same frequency and peak-to-peak variation. 

The second observation is that even if $\phi_D$ is small so that $P_{fc} << \frac{I_0}{4}$, it is still possible for $\text{DoM} > 1$. The reason why this can happen is because the DC value is affected by both $\phi_s$ and $\phi_D$. Therefore, it can be that $\text{DC} \approx 0$, leading to $\text{DoM} > 1$ even though $P_{fc} << \frac{I_0}{4}$.

Fig.~\ref{fig:s3} and Fig.~\ref{fig:s4} show the plots of how the variables that influence $I(t)$ vary as a function of the incoming angle of the plane wave ($\theta,\psi$). For Fig.~\ref{fig:s3} and Fig.~\ref{fig:s4}, $L = 505~\mu \text{m}$ ($\frac{3.7696~\text{MHz}}{3.7337~\text{MHz}} \times 500~\mu\text{m} \approx 505~\mu\text{m}$), $n_o = 2.2965$, $n_e = 2.215$. These variables are chosen to be consistent with experimental measurements in the main text. Notice that even though most points on the DoM for Fig.~\ref{fig:s4} are small, there are specific angles for which $\text{DoM} > 1$ due to the explanation in the previous paragraph.

\section{Contribution of Electro-Optic Effect}
The linear electro-optic effect can also modulate the polarization of light in lithium niobate. The modified index ellipsoid when the electric field $E$ is applied along the y direction through the surface electrodes is as follows:

\begin{gather}
\Big(\frac{1}{n_o^2} - r_{22}E\Big)x^2 + \Big(\frac{1}{n_o^2} + r_{22}E\Big)y^2 + \Big(\frac{1}{n_e^2}\Big)z^2 + 2yzr_{51}E = 1 \label{Eq.41} \tag{S41}
\end{gather}

We now apply a rotation to the yz axis such the the new form is diagonal (as was carried out for the strain tensor in Section 1). Using the coordinate transformation in equation \eqref{Eq.3}, equation \eqref{Eq.41} can be transformed into:

\begin{gather}
\Big(\frac{1}{n_o^2} - r_{22}E\Big)x^2 + \Big(\frac{1}{n_o^2} + r_{22}E + r_{51}E\text{tan}\theta\Big)y'^2 + \Big(\frac{1}{n_e^2} - r_{51}E\text{tan}\theta\Big)z'^2 = 1 \nonumber \\
\text{tan}(2\theta) = \frac{2r_{51}E}{\frac{1}{n_o^2} + r_{22}E - \frac{1}{n_e^2}}\label{Eq.42} \tag{S42}
\end{gather}

Since $\text{tan}\theta \ll 1$, we neglect the modulations of the $y'$ and $z'$ axis which include the $\text{tan}\theta$ term. The modified index ellipsoid can therefore be approximated as:

\begin{gather}
\Big(\frac{1}{n_o^2} - r_{22}E\Big)x^2 + \Big(\frac{1}{n_o^2} + r_{22}E \Big)y'^2 + \Big(\frac{1}{n_e^2}\Big)z'^2 = 1 \label{Eq.43} \tag{S43}
\end{gather}

Using the electro-optic coefficient of $r_{22} = 6.7 \times 10^{-12}$~m/V and an electric field of $E = 2 \times 10^{3}$~V/m (2Vpp applied to wafer electrodes), time-varying birefringence induced by the linear electro-optic effect is approximately $\Delta n_{eo} = 0.5 E n_o^3 r_{22} \approx 8.2 \times 10^{-8}$. For the same simulation, $\bar{S}_{yz} = 1.13 \times 10^{-5}$ with $p_{14} = 0.05$. The time-varying birefringence induced by the photoelastic effect is approximately $\Delta n_{pe} = n_o^3p_{14}\bar{S}_{yz} \approx 6.9 \times 10^{-6}$.

\begin{gather}
\frac{\Delta n_{eo}}{\Delta n_{pe}} \approx 0.01 \label{Eq.44} \tag{S44}
\end{gather}

We see that the photoelastic effect is approximately two orders of magnitude stronger than the linear electro-optic effect. Therefore, the contribution of the linear electro-optic effect can be neglected. We carry out a measurement to verify this finding. We measure the contribution of the linear electro-optic effect off resonance at 3~MHz. The measurement setup and the results are shown in Fig.~\ref{fig:s5}. It is clear that the electro-optic effect has a negligible contribution.

\begin{figure*}[t!]
\centering
\includegraphics[width=0.8\textwidth]{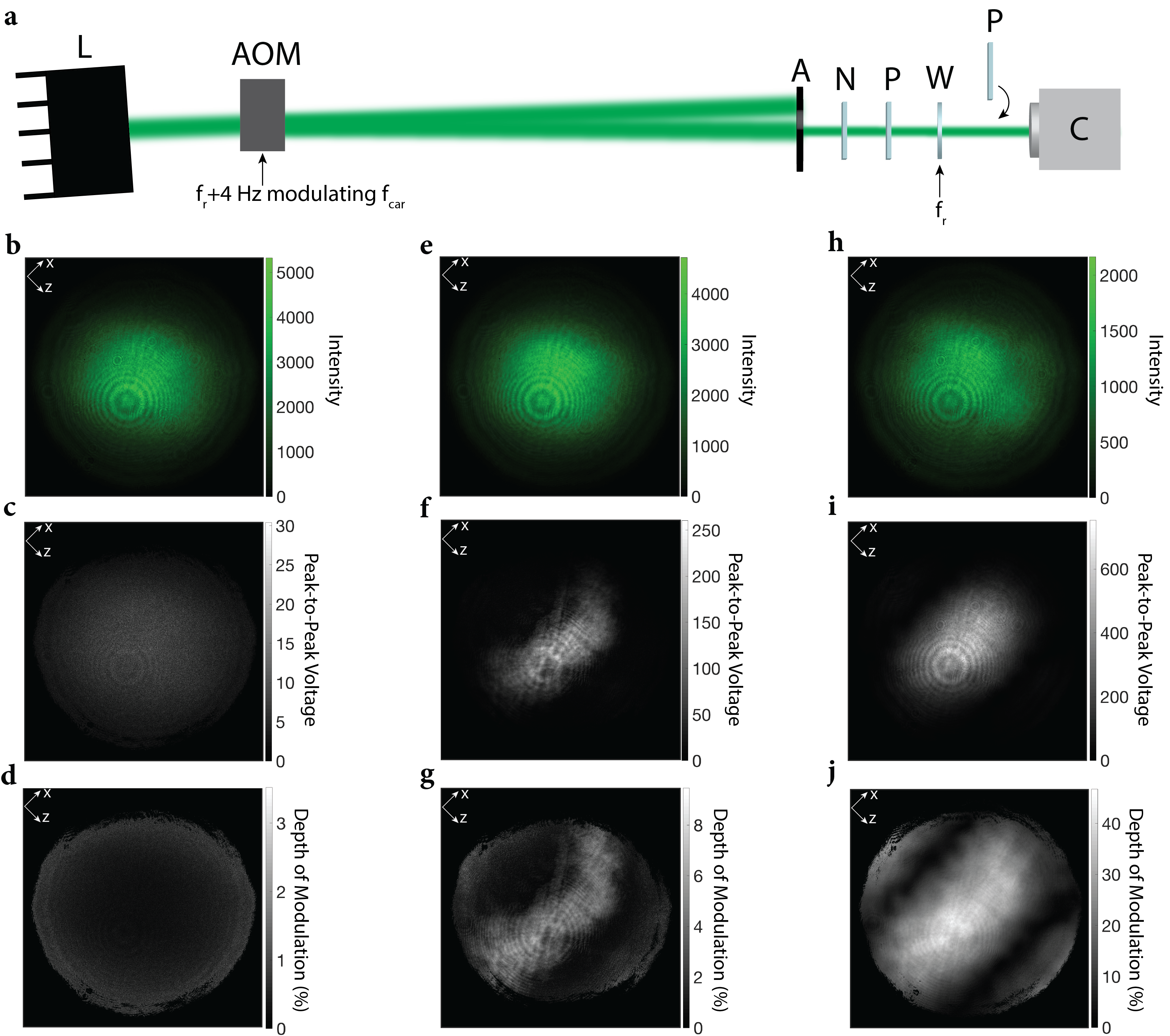}
\caption{\textbf{Measurement of modulation efficiency}. \textbf{a}, Schematic of the characterization setup is shown. The setup includes a laser (L) with a wavelength of 532~nm that passes through a free-space acousto-optic modulator (AOM). The laser beam is intensity modulated at $f_r + 4~\text{Hz} = 3.733704~\text{MHz}$ by modulating the carrier frequency $f_{car} = 80~\text{MHz}$ exciting the AOM. The setup also includes an aperture (A) with a diameter of 1~cm, a neutral density filter (N), two polarizers (P) with transmission axis $\hat{t}$, wafer (W), and a standard CMOS camera (C). The camera detects the intensity modulated laser beam with a frame rate of 30~Hz. \textbf{b}, Time-averaged intensity profile of the laser beam detected by the camera is shown for $\phi = 0$ and when no voltage is applied to wafer surface electrodes. The second polarizer (P) is inserted between (W) and (C) for this measurement. \textbf{c}, The peak-to-peak variation at 4~Hz of the laser beam is shown per pixel for $\phi = 0$ and when no voltage is applied to wafer surface electrodes. The second polarizer (P) is inserted between (W) and (C) for this measurement. \textbf{d}, The DoM at 4~Hz of the laser beam is shown per pixel for $\phi = 0$ and when no voltage is applied to wafer surface electrodes. The second polarizer (P) is inserted between (W) and (C) for this measurement. \textbf{e}, Time-averaged intensity profile of the laser beam detected by the camera is shown for $\phi = 0$ and when 90~mW of RF power at $f_r$ is applied to wafer surface electrodes. The second polarizer (P) is not inserted between (W) and (C) for this measurement. \textbf{f}, The peak-to-peak variation at 4~Hz of the laser beam is shown per pixel for $\phi = 0$ and when 90~mW of RF power at $f_r$ is applied to wafer surface electrodes. The second polarizer (P) is not inserted between (W) and (C) for this measurement. \textbf{g}, The DoM at 4~Hz of the laser beam is shown per pixel for $\phi = 0$ and when 90~mW of RF power at $f_r$ is applied to wafer surface electrodes. The second polarizer (P) is not inserted between (W) and (C) for this measurement. \textbf{h}, Time-averaged intensity profile of the laser beam detected by the camera is shown for $\phi = 0$ and when 90~mW of RF power at $f_r$ is applied to wafer surface electrodes. The second polarizer (P) is inserted between (W) and (C) for this measurement. \textbf{i}, The peak-to-peak variation at 4~Hz of the laser beam is shown per pixel for $\phi = 0$ and when 90~mW of RF power at $f_r$ is applied to wafer surface electrodes. The second polarizer (P) is inserted between (W) and (C) for this measurement. \textbf{j}, The DoM at 4~Hz of the laser beam is shown per pixel for $\phi = 0$ and when 90~mW of RF power at $f_r$ is applied to wafer surface electrodes. The second polarizer (P) is inserted between (W) and (C) for this measurement.}
\label{fig:s6}
\end{figure*}

\section{Intensity Modulation Efficiency}
In this section, we will calculate the modulation efficiency of the intensity modulator experimentally, and extract the volume average strain in the wafer. The measurement setup in Fig.~\ref{fig:s6}\textbf{a} is used. For measurements shown in Fig.~\ref{fig:s6}\textbf{b,c,d}, the second polarizer is inserted between the wafer (W) and the camera (C), but no voltage is applied to the wafer surface electrodes. This is a control measurement to make sure no modulation is present on the laser beam without voltage applied to the electrodes. 

Fig.~\ref{fig:s6}\textbf{e,f,g} show the measurements when 90~mW of RF power is applied to the wafer surface electrodes, but the second polarizer is removed. This measurement is done to show that the dominant intensity modulation effect manifests itself when the second polarizer is present (and therefore intensity modulation is caused via polarization modulation). The intensity modulation seen in Fig.~\ref{fig:s6}\textbf{f} without the second polarizer is most likely due to anisotropic Bragg diffraction~\cite{dixon1967acoustic,acousto_optic_tunable}, where some portion of the laser beam is scattered and Doppler shifted by $f_r = 3.7337~\text{MHz}$. The camera detects the beating between the optical pump beam and the scattered and Doppler shifted beam, which appears as intensity modulation.

Fig.~\ref{fig:s6}\textbf{h,i,j} show the measurements when 90~mW of RF power is applied to the wafer surface electrodes, and the second polarizer is inserted between the wafer (W) and the camera (C). It is clear that the dominant intensity modulation mechanism is polarization modulation when we compare Fig.~\ref{fig:s6}\textbf{i} to Fig.~\ref{fig:s6}\textbf{f}. Using Fig.~\ref{fig:s6}\textbf{e,h,i}, we can calculate $|\text{sin}\phi_s|$ and $\phi_D$. Without the second polarizer, the camera detects $\frac{I_0}{2}$. When the second polarizer is inserted, the DC value is: $\frac{I_0}{2}\big(c_o^4 + c_e^4 + 2c_o^2c_e^2\text{cos}\phi_sJ_0(\phi_D)\big)$. Since the laser beam is coming at a normal angle to the wafer, $c_o = c_e = \frac{1}{\sqrt{2}}$. We take the ratio of the sum of pixels in Fig.~\ref{fig:s6}\textbf{h} to Fig.~\ref{fig:s6}\textbf{e}, which yields approximately 0.47. To simplify the calculations, we assume it is 0.5. For this ratio to be 0.5, $\text{cos}\phi_s$ = 0. This means $|\text{sin}\phi_s| = 1$. The modulation power at frequency $f_r$ can now be expressed as $\frac{I_0J_1(\phi_D)}{2}$. We can estimate $J_1(\phi_D)$ by taking the ratio of the pixels in Fig.~\ref{fig:s6}\textbf{i} to Fig.~\ref{fig:s6}\textbf{e}. Since we know that $\phi_D < 1$, we use the following approximation: $J_1(\phi_D) \approx \frac{\phi_D}{2}$. The average across all pixels for $\phi_D$ is $\bar{\phi}_D \approx 0.15$. We now estimate the volume average strain in the wafer as:

\begin{gather}
\bar{S}_{yz} = \frac{\bar{\phi}_D \lambda}{2\pi L p_{14} n_o^3} \approx 4.15 \times 10^{-5}  \label{Eq.45} \tag{S45}
\end{gather}

The reason behind using the peak-to-peak variation in Fig.~\ref{fig:s6} is to compensate for the heterodyne detection. The intensity modulation imparted on the laser beam by the modulator generates a beat tone at 4~Hz, but at half the power.

\section{$\boldsymbol{s_{11}}$ Measurement at Different RF Power Levels}
\begin{figure*}[t!]
\centering
\includegraphics[width=0.8\textwidth]{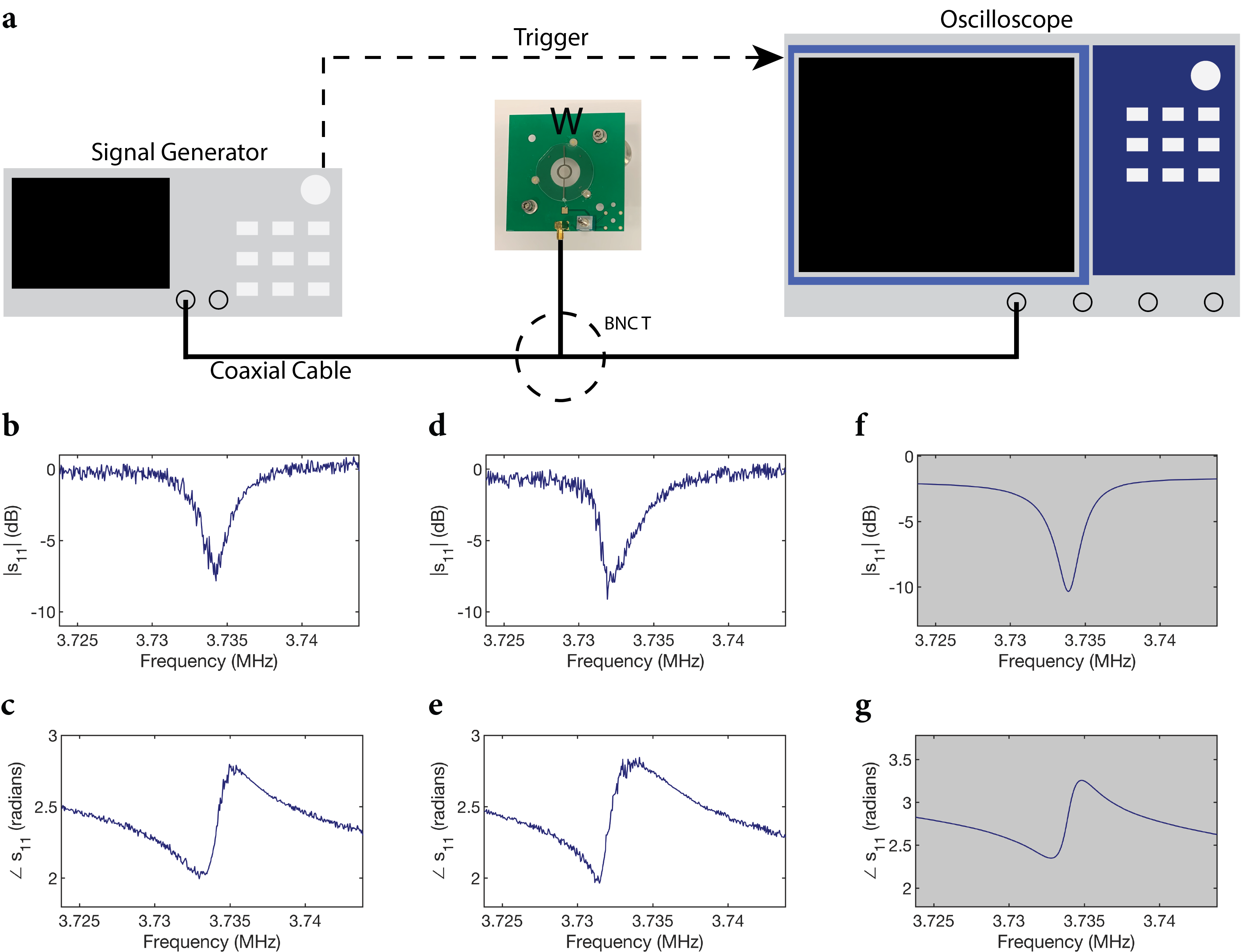}
\caption{\textbf{Measuring $\boldsymbol{s_{11}}$ at different excitation powers}. \textbf{a}, Schematic of the characterization setup is shown. The setup includes a signal generator exciting the wafer (W) surface electrodes and triggering an oscilloscope. The oscilloscope detects the voltage on the wafer surface electrodes with high input impedance ($1 ~ \text{M} \Omega$). \textbf{b}, Computed $|s_{11}|$ using the setup shown in \textbf{a} with respect to $50~ \Omega$ using 4.0~dBm excitation power with a bandwidth of 1~Hz for the desired acoustic mode. \textbf{c}, Computed $\angle s_{11}$ using the setup shown in \textbf{a} with respect to $50~ \Omega$ using 4.0~dBm excitation power with a bandwidth of 1~Hz for the desired acoustic mode. \textbf{d}, Computed $|s_{11}|$ using the setup shown in \textbf{a} with respect to $50~ \Omega$ using 19.5~dBm excitation power with a bandwidth of 1~Hz for the desired acoustic mode. \textbf{e}, Computed $\angle s_{11}$ using the setup shown in \textbf{a} with respect to $50~ \Omega$ using 19.5~dBm excitation power with a bandwidth of 1~Hz for the desired acoustic mode. \textbf{f}, $|s_{11}|$ scan using a VNA with respect to $50~ \Omega$ with 0~dBm excitation power and with a bandwidth of 20~Hz for the desired acoustic mode. \textbf{g}, $\angle s_{11}$ scan using a VNA with respect to $50~ \Omega$ with 0~dBm excitation power and with a bandwidth of 20~Hz for the desired acoustic mode.}
\label{fig:s7}
\end{figure*}

In this section, we show how the $s_{11}$ properties of the modulator change when excited at different RF power levels. The $s_{11}$ is measured with 0~dBm power using a VNA (vector network analyzer), and the magnitude and phase plots are shown in Fig.~\ref{fig:s7}\textbf{f,g}. Since optical characterization of the modulator is performed at a higher RF power (90~mW = 19.5~dBm) to achieve higher modulation efficiencies, we also need to measure the $s_{11}$ at this RF power level. The setup shown in Fig.~\ref{fig:s7}\textbf{a} is used for this measurement, where a signal generator allows us to measure at power levels higher than possible with the VNA. We see that for an excitation power of 4.0~dBm, we get a very similar response to what was measured using the VNA, as shown in Fig.~\ref{fig:s7}\textbf{b,c}. This confirms that the measurement setup is consistent with the VNA measurements. Next we perform an $s_{11}$ measurement with 19.5~dBm of RF power; the RF power used for optical characterization. We see a shift in the resonance frequency of approximately 2~kHz in Fig.~\ref{fig:s7}\textbf{d,e}. This shift is most likely caused by a change in the lithium niobate temperature due to the higher excitation power. For a frequency shift of 2~kHz, the excited volume average strain level in COMSOL shows negligible change.

\begin{figure*}[t!]
\centering
\includegraphics[width=0.8\textwidth]{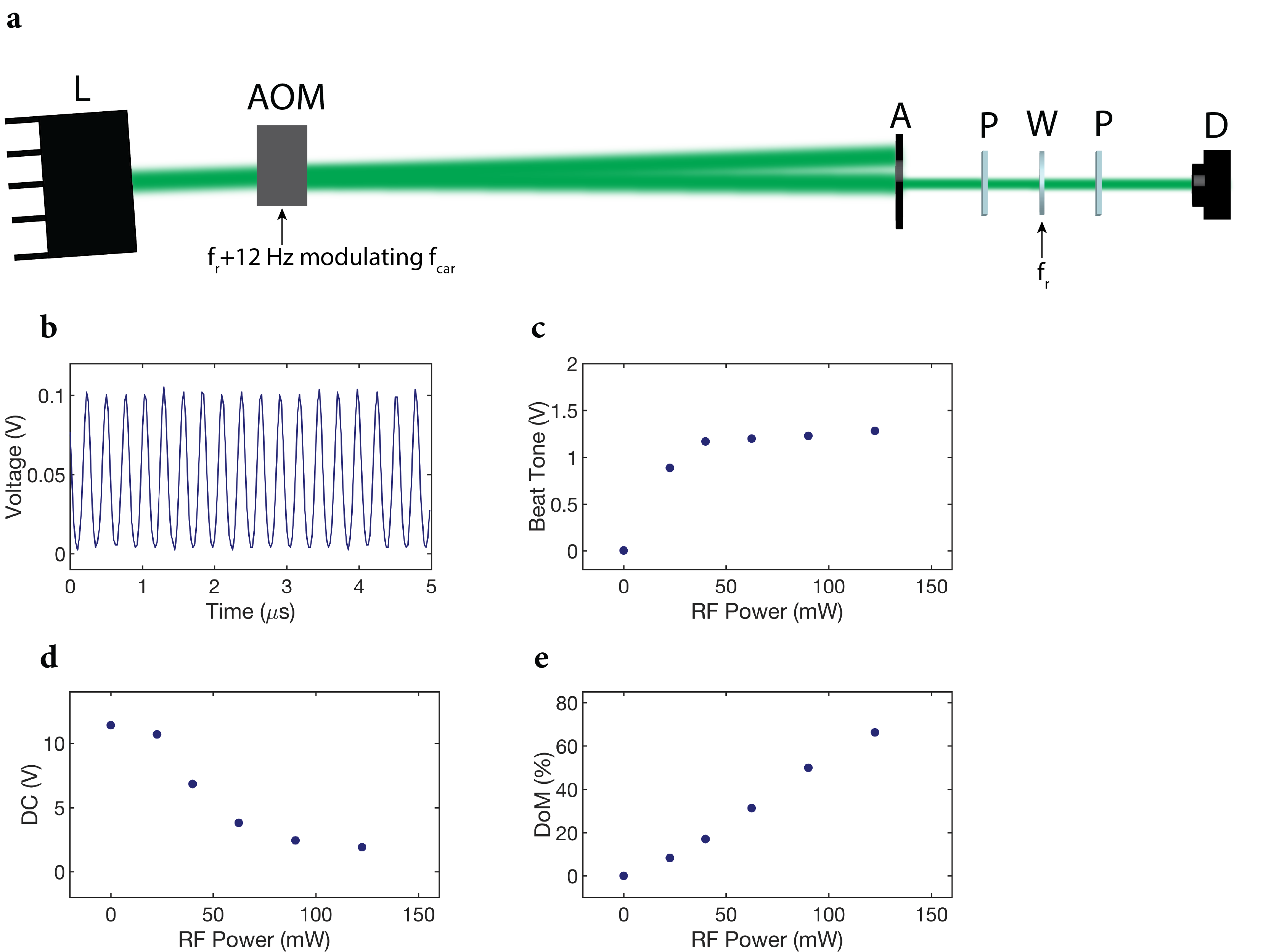}
\caption{\textbf{Depth of modulation measurement at different RF excitation power levels}. \textbf{a}, Schematic of the characterization setup is shown. The setup includes a laser (L) with a wavelength of 532~nm that passes through a free-space acousto-optic modulator (AOM). The laser beam is intensity modulated at $f_r + 12~\text{Hz} = 3.733712~\text{MHz}$ by modulating the carrier frequency $f_{car} = 80~\text{MHz}$ exciting the AOM. The setup also includes an aperture (A) with a diameter of 2.5~mm, two polarizers (P) with transmission axis $\hat{t}$, wafer (W), and a free-space photodetector (D). The photodetector detects the intensity modulated laser beam with sampling rate of 40~MHz. \textbf{b}, Measured intensity modulation profile for the laser beam using the photodetector is shown when no voltage is applied to wafer surface electrodes.  \textbf{c}, The peak-to-peak variation for the beat tone at 12~Hz is shown when the wafer surface electrodes are excited at different RF power levels. \textbf{d}, The DC level for the intensity of the laser beam is shown when the wafer surface electrodes are excited at different RF power levels. \textbf{e}, The DoM at 12~Hz for the laser beam is shown when the wafer surface electrodes are excited at different RF power levels.}
\label{fig:s8}
\end{figure*}

\section{Intensity Modulation Efficiency of Different Modulation Mechanisms}
In this section, we will calculate the modulation efficiency of the proposed intensity modulator and other modulators relying on different modulation mechanisms. We first begin by calculating the modulation efficiency for the approach discussed in this paper: resonant photoelastic modulation. The modulator has a thickness of 0.5~mm and a usable aperture diameter of 1~cm. The volume average dynamic phase is $\bar{\phi}_D = 0.15$ for RF excitation power of 90~mW. For the first order Bessel function of the first kind to be equal to 0.5, the volume average dynamic phase should be 1.2. We can now calculate the modulation efficiency by computing the required RF power to switch the photoelastic modulator at 3.7~MHz, at a wavelength of 532~nm, and with an input aperture of $1~\text{cm}^2$:

\begin{gather}
\text{P}_{on*} = \Big(\frac{1~\text{cm}^2}{\pi*0.5^2~\text{cm}^2}\Big) \times \Big(\frac{1.2}{0.15}\Big)^2 \times 0.09~\text{W} \approx 7.4~\text{W} \label{Eq.46} \tag{S46}
\end{gather}

We now calculate for a commercial transverse resonant Pockels cell~\cite{resonant_transverse_Pockels}. The Pockels cell has a thickness of 57.2~mm and an aperture diameter of 2~mm. The half-wave voltage at 633~nm is 15~V. To make a fair comparison to our modulator and have the same temperature tolerance, we assume a thickness of 0.5~mm for the Pockels cell. We can now calculate the modulation efficiency by computing the required RF power to switch the Pockels cell at 3.7~MHz, at a wavelength of 532~nm, and with an input aperture of $1~\text{cm}^2$ (3~dB cutoff frequency due to RC time constant is not taken into account):

\begin{gather}
\text{P}_{on1} = \frac{\Big(\frac{15~\text{V}}{\sqrt{2}} \times \frac{532~\text{nm}}{633~\text{nm}} \times \frac{11.3~\text{mm}}{2~\text{mm}} \times \frac{57.2~\text{mm}}{0.5~\text{mm}}\Big)^2}{50~\Omega} \approx 6.7 \times 10^5~\text{W} \label{Eq.47} \tag{S47}
\end{gather}

We now calculate for a commercial longitudinal Pockels cell~\cite{longitudinal_Pockels}. The Pockels cell has an aperture diameter of 9~mm and a thickness of 36~mm. The capacitance is 8~pF and the half-wave voltage is 3.3~kV. To make a fair comparison to our modulator and have the same temperature tolerance, we assume a thickness of 0.5~mm for the Pockels cell. We can now calculate the modulation efficiency by computing the required RF power to switch the Pockels cell at 3.7~MHz, at a wavelength of 532~nm, and with an input aperture of $1~\text{cm}^2$ (3~dB cutoff frequency due to RC time constant is not taken into account): 

\begin{gather}
\text{P}_{on2} = \frac{1}{2}\Big(\frac{11.3~\text{mm}}{9.0~\text{mm}}\Big)^2 \times \Big(3,300~\text{V}\Big)^2 \times \frac{8 \times 10^{-12}~\text{F} \times 36~\text{mm}}{0.5~\text{mm}} \times 3.7 \times 10^6~\text{Hz} \approx 1.83 \times 10^4~\text{W} \label{Eq.48} \tag{S48}
\end{gather}

We now calculate for an electroabsorption modulator relying on the quantum-confined Stark effect~\cite{electroabsorption_patent} for modulating the intensity of a laser beam. The modulator has aperture dimensions of 6~mm by 7~mm. The power consumption is 14~W, yielding a depth of modulation of approximately 60\%. The wavelength of operation is 860~nm. We can now calculate the modulation efficiency by computing the required RF power to switch the modulator at 3.7~MHz, with an input aperture of $1~\text{cm}^2$ and assuming linear relationship between applied electric field and depth of intensity modulation (3~dB cutoff frequency due to RC time constant is not taken into account): 

\begin{gather}
\text{P}_{on3} = 14~\text{W} \times \frac{3.7~\text{MHz}}{10~\text{MHz}} \times \Big(\frac{100\%}{60\%}\Big)^2 \times \Big(\frac{1~\text{cm}^2}{0.6~\text{cm} \times 0.7~\text{cm}}\Big) \approx  34~\text{W} \label{Eq.49} \tag{S49}
\end{gather}

We now calculate for a gate-tunable metasurface~\cite{gate_tunable_metasurface}. For this modulator, the unit area capacitance is $14~\text{fF}/\mu\text{m}^2$, and 15\% amplitude modulation is observed for an applied voltage of 2~V. The reported wavelength of operation is 1,550~nm. We can now calculate the modulation efficiency by computing the required RF power to switch the modulator at 3.7~MHz, with an input aperture of $1~\text{cm}^2$ assuming linear relationship between applied electric field and intensity modulation efficiency (3~dB cutoff frequency due to RC time constant is not taken into account): 

\begin{gather}
\text{P}_{on4} = \Big(\frac{100\%}{15\%}\Big)^2 \times \Big(\frac{1}{2} \times 14 \times 10^{-15}~\text{F} \times (2~\text{V})^2 \times 3.7 \times 10^6~\text{Hz}\Big) \times \Big(\frac{1~\text{cm}}{1~\mu \text{m}}\Big)^2  \approx  460~\text{W} \label{Eq.50} \tag{S50}
\end{gather}

We now calculate for a plasmonic nanoresonator~\cite{slm_steering_brongersma}. For this modulator, the switching energy is $283~\text{fJ}/\mu\text{m}^2$, and the resulting amplitude modulation is 48\%. The reported wavelength of operation is $1.34~\mu\text{m}$. We can now calculate the modulation efficiency by computing the required RF power to switch the modulator at 3.7~MHz, with an input aperture of $1~\text{cm}^2$ assuming linear relationship between applied electric field and intensity modulation efficiency (3~dB cutoff frequency due to RC time constant is not taken into account): 

\begin{gather}
\text{P}_{on5} = 283 \times 10^{-15}~\text{J} \times \Big(\frac{1~\text{cm}}{1~\mu \text{m}}\Big)^2 \times 3.7 \times 10^6~\text{Hz} \times \Big(\frac{100\%}{48\%}\Big)^2 \approx  454~\text{W} \label{Eq.51} \tag{S51}
\end{gather}

\end{document}